\documentclass[a4paper,11pt]{article}
\pdfoutput=1

\usepackage{jcappub}
\usepackage[T1]{fontenc}
\usepackage{physics}
\usepackage{longtable}
\usepackage{booktabs}
\usepackage{verbatim}

\newcommand{\Ito}{It\^{o}}

\newcommand{\MPl}{M_\mathrm{Pl}}

\newcommand{\cond}{\mspace{1mu}|\mspace{1mu}}

\newcommand{\PFPT}{P_\text{FPT}}

\newcommand{\R}{\mathcal{R}}
\newcommand{\PR}{\mathcal{P}_\R}

\newcommand{\iFi}{{}_1 F_1}

\newcommand{\paperI}{\emph{Paper I}}
\newcommand{\LFP}[1]{\mathcal{L}_{\text{FP},#1}}
\newcommand{\adLFP}[1]{\mathcal{L}^\dagger_{\text{FP},#1}}
\newcommand{\saL}[1]{\mathcal{L}_{\text{s-a},#1}}
\newcommand{\tphi}{\tilde{\phi}}

\newcommand{\varlam}{\tilde{\lambda}}
\newcommand{\varw}{\tilde{w}}
\newcommand{\varN}{\tilde{N}}
\newcommand{\expvarN}{\langle \varN \rangle}

\title{Stochastic constant-roll inflation beyond the hilltop with the spectral method}
\author{Eemeli Tomberg}
\affiliation{Cosmology, Universe and Relativity at Louvain (CURL), Institute of Mathematics and Physics, University of Louvain, 2 Chemin du Cyclotron, 1348 Louvain-la-Neuve, Belgium}
\emailAdd{eemeli.tomberg@uclouvain.be}

\abstract{
Stochastic inflation can be used to study large inflationary perturbations. This paper presents such a study for a quadratic hilltop potential, corresponding to constant-roll inflation. I solve the perturbation distribution using the spectral method, with detailed solutions of the eigenvalues and eigenfunctions of the Fokker--Planck operator. Contrary to previous studies of stochastic constant-roll inflation, the solution allows trajectories that cross the hilltop and get stuck near a reflecting boundary on the other side, tunneling out slowly in a way dictated by the lowest eigensolution. Despite their rarity, these trajectories turn out to dominate the mean first-passage time. For this reason, I argue the mean does not properly describe the inflationary background. Using the median instead, I compute the distribution of the coarse-grained $\Delta N$ distribution and show that its well-known exponential tail first flattens out and then forms a peak near a maximal $\Delta N$ value. I argue similar intricacies arise in primordial black hole models.
}

\begin{document}

\maketitle

\section{Introduction}
\label{sec:intro}
Cosmic inflation \cite{Starobinsky:1980te,Kazanas:1980tx,Sato:1981qmu,Guth:1980zm} explains the homogeneity and isotropy of the large-scale Universe, while also predicting fluctuations around this background. Typical fluctuations, such as those seen in the cosmic microwave background radiation \cite{Planck:2018jri, ACT:2025fju}, are small and well-described by linear perturbation theory. Large fluctuations may be relevant for eternal inflation \cite{Winitzki:2008zz} and the formation of primordial black holes (PBHs) \cite{Hawking:1971ei, Carr:1975qj, Green:2020jor, Carr:2025kdk}. They can be studied with \emph{stochastic inflation}.

In stochastic inflation, we divide the physics into short and long wavelengths. The short wavelengths approximately follow linear perturbation theory, while the long wavelengths describe a set of independently evolving FLRW universes, one in each Hubble patch, as per the separate universe approximation \cite{Salopek:1990jq, Wands:2000dp}. As space expands, short-wavelength Fourier modes drift over the coarse-graining boundary, contributing stochastic kicks to the local FLRW evolution. As a result, the local inflaton field (and other background quantities) undergo stochastic motion \cite{Starobinsky:1986fx}, making the inflationary number of e-folds $N$ a random variable. The $\Delta N$ formalism \cite{Sasaki:1995aw, Sasaki:1998ug, Wands:2000dp, Lyth:2004gb} translates $N$ into the coarse-grained curvature perturbation $\zeta$, the conventional perturbation variable. Since the local FLRW evolution is non-perturbative, the stochastic $\Delta N$ formalism \cite{Fujita:2013cna, Fujita:2014tja, Vennin:2015hra} can handle large $\zeta$ values beyond the validity of linear perturbation theory, at the cost of lost (presumably small) gradient effects near the Hubble scale.

Traditionally, stochastic inflation is applied to slow-roll models \cite{Vennin:2020kng, Vennin:2024yzl}. However, stochastic inflation computations have developed considerably in recent years \cite{PerreaultLevasseur:2013kfq, Vennin:2015hra, Pattison:2017mbe, Firouzjahi:2018vet, Biagetti:2018pjj, Ezquiaga:2018gbw, Cruces:2018cvq, Prokopec:2019srf, Pattison:2019hef, Ezquiaga:2019ftu, Rigopoulos:2021nhv, Ahmadi:2022lsm, Pattison:2021oen, Artigas:2021zdk, Tada:2021zzj, Cruces:2021iwq, Animali:2022otk, Rigopoulos:2022gso, Mahbub:2022osb, Jackson:2022unc, Cruces:2022imf, Briaud:2023eae, Mishra:2023lhe, Jackson:2024aoo, Sharma:2024fbr, Cruces:2024pni, Launay:2024qsm, Mizuguchi:2024kbl, Blachier:2025iwk, Blachier:2025tcq, Briaud:2025ayt, Choudhury:2025kxg, Launay:2025lnc, Prokopec:2025uvz, Cruces:2026yvs, Kawasaki:2026hnx, Animali:2026omc}, especially in the context of PBH models, which can break slow-roll. A particularly interesting example is that of a quadratic hilltop potential, where the inflaton undergoes \emph{constant-roll} (CR) inflation: the first slow-roll parameter $\epsilon_1$ is small, while the second slow-roll parameter $\epsilon_2$ is a potentially large positive constant \cite{Motohashi:2014ppa, Karam:2022nym}. The attractor nature of constant roll allows a simple stochastic formulation akin to slow roll \cite{Tomberg:2022mkt}, and the resulting $\zeta$ distribution was solved by the current author in reference \cite{Tomberg:2023kli}, which I henceforth call \paperI. The results of \paperI\ apply for stochastic trajectories on the `classical' side of the hilltop where inflation ends at an absorbing boundary. In this paper, I develop the solution further by including trajectories that cross beyond the hilltop to a new diffusion-dominated regime. The hilltop-crossing trajectories cluster around a reflecting boundary placed in this regime. Due to the stochastic kicks, they slowly trickle back out, ending inflation at a late time.

In the stochastic formulation, the inflaton field and inflation duration statistics can be solved from the \emph{Fokker--Planck equation} and its adjoint. These equations can be solved via a spectral decomposition based on the eigenvalues and eigenfunctions of the Fokker--Planck operator. The spectral method is a standard technique for solving separable linear partial differential equations, but it is underutilized in modern stochastic inflation studies.\footnote{Many studies use an alternative method based on the system's \emph{characteristic function}. I compare the two methods in Section~\ref{sec:discussion}.} It is well-suited for describing the long-time behavior of the system, associated with the lowest eigensolutions and large $\zeta$ perturbations. In this paper, I present the technique in detail and apply it to the constant-roll system described above.\footnote{The spectral decomposition of constant-roll hilltop inflation was studied earlier in \cite{Tomberg:2025fku} with two absorbing boundaries and an emphasis on the lowest eigensolution; this paper complements and expands the work of \cite{Tomberg:2025fku}. The spectral method was also recently studied in \cite{Mishra:2026xal}.}

Solving the tail of the $\zeta$ distribution turns out to be challenging: the $\Delta N$ formalism is based on comparing the local expansion to the background value, and the background is ambiguous in the presence of the hilltop-crossing trajectories. Typically, the background is defined as an average over the stochastic solutions. However, in the constant-roll system, different average measures differ significantly from each other. I argue that the traditionally-used mean $N$ is not a good description for the background when rare, extreme perturbations are present, and advocate for using the median $N$ instead. Adopting this convention, the $\zeta$ distribution exhibits the well-known exponential tail \cite{Pattison:2017mbe, Ezquiaga:2019ftu, Tomberg:2023kli} at moderate $\zeta$ values; however, for larger $\zeta$, the tail first flattens out and then forms a sharp peak near a maximal $\Delta N$ value. This novel behavior is the paper's main result. It arises from the diffusion regime and can't be reproduced by classical $\Delta N$ techniques.

The paper is organized as follows. In Section~\ref{sec:stoch_inflation}, I discuss solving the Fokker--Planck equation using the spectral method. I discuss the spectral decomposition of the constant-roll hilltop model in various limits in Section~\ref{sec:cr} and use it to solve the distributions of the inflaton field and the curvature perturbation $\zeta$ in Sections~\ref{sec:time_evolution_of_P} and~\ref{sec:FPT_statistics}. I compare the results to the literature in Section~\ref{sec:discussion} and conclude in Section~\ref{sec:conclusions}.

Throughout the paper, I use natural units with $\MPl=\hbar=c=1$, where $\MPl$ is the reduced Planck mass, $\hbar$ is the reduced Planck constant, and $c$ is the speed of light.

\section{Stochastic inflation with the spectral method}
\label{sec:stoch_inflation}
In stochastic inflation, the inflaton field $\phi$ follows the Langevin equation (see, e.g., \cite{Vennin:2020kng, Vennin:2024yzl} for this and other basic equations presented in Section~\ref{sec:FP})
\begin{equation} \label{eq:Langevin}
    \frac{\dd \phi}{\dd N} = \mu(\phi) + \sigma(\phi)\xi(N) \, ,
\end{equation}
where $\mu$ is the classical drift, $\sigma$ is the diffusion coefficient, $\xi$ is white Gaussian noise with the correlator $\expval{\xi(N)\xi(N')}=\delta(N-N')$, and $N$ is the number of e-folds of spatial expansion, the correct time variable for the stochastic formalism \cite{Finelli:2008zg, Vennin:2015hra}. In the usual case of slow-roll inflation, $\mu(\phi)=V'(\phi)/V(\phi)$ and $\sigma(\phi)=H(\phi)/(2\pi)$, where $V$ is the inflaton potential and $H$ is the Hubble parameter, related to each other by the Friedmann equation $3H^2(\phi)=V(\phi)$. I will keep the $\mu$ and $\sigma$ functions generic for now, specializing in the modified case of constant-roll inflation in Section~\ref{sec:cr}.

\subsection{Fokker--Planck equation}
\label{sec:FP}
Let us start by reviewing standard results for the field statistics.
As $\phi$ evolves stochastically, we can track its probability distribution $P(\phi, N)$ in time. The distribution follows the \emph{Fokker--Planck equation}
\begin{equation} \label{eq:FP}
    \partial_N P(\phi, N) = \LFP{\phi}P(\phi, N) \equiv \partial_\phi \qty[\partial_\phi \qty(\frac{1}{2}\sigma^2(\phi)P(\phi, N)) - \mu(\phi)P(\phi, N) ] \, ,
\end{equation}
which can be derived from the Langevin equation \eqref{eq:Langevin} (by adopting the \Ito\ convention; see, e.g., \cite{Tomberg:2024evi}). The equation defines the \emph{Fokker--Planck operator} $\LFP{\phi}$, a differential operator in $\phi$ and the generator of time translations in $P(\phi,N)$.

To characterize the time evolution, let us define the \emph{probability current}
\begin{equation} \label{eq:j}
    j(\phi, N) \equiv -\partial_\phi \qty(\frac{1}{2}\sigma^2(\phi)P(\phi, N)) + \mu(\phi)P(\phi, N) 
\end{equation}
so that $\LFP{\phi}P(\phi, N) = -\partial_\phi j(\phi, N)$, and the \emph{survival probability}
\begin{equation} \label{eq:S}
    S(N, \phi_1, \phi_2) \equiv \int_{\phi_1}^{\phi_2} \dd \phi \ P(\phi, N) \quad \text{with} \quad \phi_1 < \phi_2 \, .
\end{equation}
Here $S(N,\phi_1,\phi_2)$ gives the probability to find $\phi$ between $\phi_1$ and $\phi_2$ at time $N$. By the Fokker--Planck equation, the time derivative of $S$ reads
\begin{equation} \label{eq:S_derivative}
    \partial_N S(N, \phi_1, \phi_2) = j(\phi_1, N) - j(\phi_2, N) \, .
\end{equation}
This shows that $j(\phi,N)$ is the flux of probability through $\phi$ (in the direction of increasing $\phi$) at $N$.

\paragraph{Boundary conditions.}
We typically consider evolution inside a finite field interval $[\phi_\text{low},\phi_\text{high}]$, fixing the initial field value $\phi=\phi_0 \in [\phi_\text{low},\phi_\text{high}]$ at an initial time $N=N_0$. We consider \emph{reflecting} and \emph{absorbing} boundaries at $\phi_\text{low}$ and $\phi_\text{high}$, corresponding to the boundary conditions
\begin{equation} \label{eq:FP_boundaries}
\begin{aligned}
    j(\phi_r, N) &= 0 \quad  (\text{reflecting boundary at } \phi_r) \, , \\
    P(\phi_a, N) &= 0 \quad (\text{absorbing boundary at } \phi_a) \, ,
\end{aligned}
\end{equation}
enforced for all $N$.
Physically, a reflecting boundary corresponds to a high classical drift just outside of $\phi_r$, pushing the field back inside. Since there is no flux through, all trajectories that hit the boundary bounce back. An absorbing boundary also corresponds to a high drift, but this time pushing the trajectories further out after they cross $\phi_a$, so that they can't cross back\footnote{Modeling the absorbing and reflecting boundaries as high drift becomes exact when $\mu(\phi) \to \pm\infty$ just beyond the boundary. In practice, absorbing and reflecting boundaries can be considered good approximations of the real physics if, beyond them, $\mu(\phi) \gg \sigma(\phi)$. In slow-roll inflation, this reduces to the well-known classicality condition $\frac{H^2}{8\pi^2\epsilon_V} \ll 1$, where $\epsilon_V$ is the first slow-roll parameter and the l.h.s. is the slow-roll power spectrum \cite{Linde:2015edk, Creminelli:2008es, Tomberg:2025fku}.}. The high drift forces the distribution $P$ to zero, so that the flux $j$ can stay finite. It is customary to place an absorbing boundary at the end-of-inflation field value: once inflation ends, it can't restart, since the stochastic motion stops. 

\paragraph{Adjoint equation.}
Let us include the initial condition in our notation as $P(\phi, N \cond \text{init.})$. In particular, $P(\phi, N \cond \phi_0, N_0)$ denotes the probability distribution of $\phi$ at time $N$ when starting from $\phi_0$ at time $N_0$. Then, clearly, we can decompose a general $P$ as follows:
\begin{equation} \label{eq:P_integral_decomposition}
    P(\phi,N \cond \text{init.})
    = \int_{\phi_\text{low}}^{\phi_\text{high}} \dd \tphi \, P(\phi, N \cond \tphi, N_1) P(\tphi, N_1 \cond \text{init.}) \, ,
\end{equation}
where $N_1$ is an arbitrary intermediate time, $0<N_1<N$, and the $\tphi$ integral represents all possible values the field can take at this time. Taking a derivative with respect to $N_1$, using \eqref{eq:FP}, and integrating by parts repeatedly, we get
\begin{equation} \label{eq:P_decomposition_derivative}
\begin{aligned}
    0
    &= \int_{\phi_\text{low}}^{\phi_\text{high}} \dd \tphi \qty[
    \partial_{N_1}P(\phi, N \cond \tphi, N_1) P(\tphi, N_1 \cond \text{init.})
    +
    P(\phi, N \cond \tphi, N_1) \LFP{\tphi} P(\tphi, N_1 \cond \text{init.})
    ] \\
    & 
    \begin{aligned}
    =\int_{\phi_\text{low}}^{\phi_\text{high}} \dd \tphi & \qty[
    \partial_{N_1}P(\phi, N \cond \tphi, N_1) +
    \frac{1}{2}\sigma^2(\tphi)\partial_{\tphi}^2 P(\phi, N \cond \tphi, N_1)
    + \mu(\tphi)\partial_{\tphi} P(\phi, N \cond \tphi, N_1)
    ] \\
    &  \times P(\tphi, N_1 \cond \text{init.})
    \end{aligned} \\
    &\phantom{=} - \qty[P(\phi, N \cond \tphi, N_1) j(\tphi, N_1 \cond \text{init.})
    + \frac{1}{2}\sigma^2(\tphi)\qty[\partial_{\tphi}P(\phi, N \cond \tphi, N_1)]P(\tphi, N_1 \cond \text{init.})]_{\phi_\text{low}}^{\phi_\text{high}} \, .
\end{aligned}
\end{equation}
Since $P(\phi, N \cond \text{init.})$ is essentially a free function (set by the initial conditions), the integrand and the boundary terms must all vanish independently. For the integrand, this gives the \emph{adjoint Fokker--Planck equation}
\begin{equation} \label{eq:adjoint_FP}
\begin{aligned}
    -&\partial_{N_1}P(\phi,N\cond \tphi, N_1) = 
    \partial_N P(\phi, N \cond \tphi, N_1) \\
    &=
    \adLFP{\tphi} 
    \partial_N P(\phi, N \cond \tphi, N_1) \equiv \frac{1}{2}\sigma^2(\tphi)\partial_{\tphi}^2 P(\phi,N\cond \tphi, N_1)
    + \mu(\tphi)\partial_{\tphi}P(\phi,N\cond \tphi, N_1) \, ,
\end{aligned}
\end{equation}
where we used the fact that, in the absence of explicit time dependence, the system can only depend on the difference $N - N_1$, and thus $\partial_{N_1} = -\partial_{N}$. Here $\adLFP{\phi}$ is the \emph{adjoint Fokker--Planck operator}\footnote{\label{ftn:adjoints} The operators $\LFP{\phi}$ and $\adLFP{\phi}$ are adjoint with respect to the inner product $f(\phi) \cdot g(\phi) \equiv \int \dd \phi f(\phi) g(\phi)$, that is, $f(\phi)\cdot \LFP{\phi}g(\phi) = \int \dd \phi f(\phi) \LFP{\phi}g(\phi) = \int \dd \phi \adLFP{\phi}f(\phi) g(\phi) = \adLFP{\phi} f(\phi)\cdot g(\phi)$, up to boundary terms.}. The boundary terms translate the boundary conditions in the evolving field $\phi$ into boundary conditions in the initial field $\tphi$ (equivalently, boundary conditions for functions obeying the Fokker--Planck equation to boundary conditions of functions obeying the adjoint Fokker--Planck equation); for the absorbing and reflecting boundaries, substituting \eqref{eq:FP_boundaries} in \eqref{eq:P_decomposition_derivative} yields
\begin{equation} \label{eq:adj_FP_boundaries}
\begin{aligned}
    \partial_{\phi_r} P(\phi, N \cond \phi_r, N_0) &= 0 \quad  (\text{reflecting boundary at } \phi_r) \, , \\
    P(\phi, N \cond \phi_a, N_0) &= 0 \quad (\text{absorbing boundary at } \phi_a) \, ,
\end{aligned}
\end{equation}
for all $\phi$ and $N$.

\paragraph{First-passage times.} The adjoint Fokker--Planck equation is useful for computing \emph{first-passage times} (FPTs). 
The first-passage-time distribution $\PFPT(N, \phi \cond \phi_0, N_0)$ gives the probability density of the time $N$ the field first crosses $\phi$, given an initial condition $\phi_0$ at $N_0$. The distribution can be obtained by setting an absorbing boundary at $\phi=\phi_a$, solving the Fokker--Planck equation, and computing the probability current through $\phi_a$:
\begin{equation} \label{eq:PFPT}
    \PFPT(N,\phi_a \cond \phi_0, N_0) = \pm j(\phi_a, N \cond \phi_0, N_0) \, .
\end{equation}
The plus (minus) sign applies for motion starting from $\phi_0 < \phi_a$ ($\phi_0 > \phi_a$).
The current correctly gives the flux through $\phi_a$, and the absorbing boundary condition ensures each stochastic trajectory only contributes its first crossing. Note that at the absorbing boundary, the flux \eqref{eq:j} simplifies to $j(\phi_a, N \cond \phi_0, N) = -\frac{1}{2}\sigma^2(\phi_a) \partial_{\phi_a} P(\phi_a, N \cond \phi_0, N)$ by \eqref{eq:adj_FP_boundaries}.

Note also that $j(\phi_a, N \cond \phi_0, N_0)$ is related to $P(\phi_a, N \cond \phi_0, N_0)$ by a linear differential operator that depends only on $\phi_a$. This operator commutes with $\adLFP{\phi_0}$, so $j$ satisfies the adjoint Fokker--Planck equation $\partial_N j(\phi_a, N \cond \phi_0, N_0) = \adLFP{\phi_0} j(\phi_a, N \cond \phi_0, N_0)$; hence, we have
\begin{equation} \label{eq:PFPT_eq}
    \partial_N \PFPT(N, \phi_a \cond \phi_0, N_0) = \adLFP{\phi_0}\PFPT(N, \phi_a \cond \phi_0, N_0)
\end{equation}
as a self-contained equation for $\PFPT$, given the boundary condition (see \eqref{eq:adj_FP_boundaries})
\begin{equation} \label{eq:PFPT_boundaries}
    \PFPT(N,\phi_a \cond \phi_a, N_0) = 0 \quad \text{for all } N \, ,
\end{equation}
together with another boundary condition inherited from $P$ (in particular, for a reflecting boundary, $\partial_{\phi_r} \PFPT(N,\phi_a \cond \phi_r, N_0) = 0$).

Below, the location of the absorbing boundary $\phi_a$ is usually clear from the setup, and we set $N_0=0$; we can then use the short-hand notation $\PFPT(N,\phi_a \cond \phi_0, N_0=0) \equiv \PFPT(N,\phi_0)$.

\subsection{Spectral method}
\label{sec:spectral_decomposition}

We wish to study the late-time behavior of quantities such as $P(\phi,N)$ and $\PFPT(N,\phi_0)$. For this purpose, we will find the spectral decomposition of the Fokker--Planck operator and its adjoint. 

Equation \eqref{eq:FP} is separable, so we seek solutions of the form
\begin{equation} \label{eq:separated_FP_solution}
    P(\phi, N) = a(N)u(\phi) \, ,
\end{equation}
for which \eqref{eq:FP} becomes
\begin{equation} \label{eq:separated_FP_equations}
\begin{gathered}
    \frac{\partial_N a(N)}{a(N)} = -\lambda = \frac{\LFP{\phi}u(\phi)}{u(\phi)} \\
    \implies a(N) \propto e^{-\lambda N} \, , \quad
    \LFP{\phi}u(\phi) = -\lambda u(\phi) \, ,
\end{gathered}
\end{equation}
where $\lambda$ is a constant; $u(\phi)$ and $\lambda$ are an eigenfunction and an eigenvalue of the operator $\LFP{\phi}$. To cast the eigenvalue problem into a familiar form, we define the new operator\footnote{When operating with these operators on a function, the multiplication with the $\phi$-dependent right-hand factors happens before the differentiation w.r.t. $\phi$ in the $\mathcal{L}$-operators.}
\begin{equation} \label{eq:self_adjoint_L}
\begin{gathered}
    \saL{\phi} \equiv \sqrt{w(\phi)}\LFP{\phi}\frac{1}{\sqrt{w(\phi)}} = \frac{1}{\sqrt{w(\phi)}}\adLFP{\phi}\sqrt{w(\phi)} \, , \\
    w(\phi) \equiv \sigma^2(\phi)\exp(-\int^\phi \frac{2\mu(\tphi)}{\sigma^2(\tphi)}\dd \tphi) \, .
\end{gathered}
\end{equation}
The relation between $\LFP{\phi}$ and $\adLFP{\phi}$ above can be verified with simple algebra. It follows that $\saL{\phi}$ is self-adjoint (in the sense described in footnote~\ref{ftn:adjoints}) and it clearly shares eigenvalues with both $\LFP{\phi}$ and $\adLFP{\phi}$, with related eigenfunctions:
\begin{equation} \label{eq:all_eigensolutions}
\begin{gathered}
    \saL{\phi} v_n(\phi) = -\lambda_n v_n(\phi)
    \iff
    \LFP{\phi} u_n(\phi) = -\lambda_n u_n(\phi)
    \iff
    \adLFP{\phi} \bar{u}_n(\phi) = -\lambda_n \bar{u}_n(\phi) \\
    \text{for} \quad
    v_n(\phi) = \sqrt{w(\phi)}u_n(\phi) = \frac{1}{\sqrt{w(\phi)}}\bar{u}_n(\phi) \, .
\end{gathered}
\end{equation}

Since $\saL{\phi}$ is self-adjoint, its eigensolutions (and thus those of $\LFP{\phi}$ and $\adLFP{\phi}$) behave analogously to, say, those of the one-dimensional Schrödinger equation; in particular,
\begin{itemize}
    \item there is a discrete set of eigenvalues, $\lambda_n$, $n=1,2,3,\dots$, with $\lambda_1 < \lambda_2 < \lambda_3 < \dots$,
    \item the $n$th eigenfunctions $u_n(\phi)$, $\bar{u}_n(\phi)$, $v_n(\phi)$ cross zero exactly $n-1$ times in the interval $]\phi_\text{low}, \phi_\text{high}[$,
    \item the eigenfunctions form an orthonormal basis (when properly normalized), with
\begin{equation} \label{eq:orthogonality}
    \int_{\phi_\text{low}}^{\phi_\text{high}} v_i(\phi) v_j(\phi) \dd \phi
    = \int_{\phi_\text{low}}^{\phi_\text{high}} \bar{u}_i(\phi) u_j(\phi) \dd \phi
    = \delta_{ij} \, ,
\end{equation}
\begin{equation} \label{eq:completeness}
    \sum_{n=1}^\infty v_n(\tilde{\phi}) v_n(\phi)
    =\sum_{n=1}^\infty \bar{u}_n(\tilde{\phi}) u_n(\phi) = \delta(\tilde{\phi} - \phi) \, ,
\end{equation}
\end{itemize}
where $\delta$ is of the Kronecker variety in \eqref{eq:orthogonality} and of the Dirac variety in \eqref{eq:completeness}.
Mathematically, these properties are guaranteed by Sturm--Liouville theory\footnote{Typically, Sturm--Liouville theory is defined in terms of a self-adjoint operator of the form $\mathcal{L}_{\text{SL},\phi} = \sqrt{w(\phi)}\saL{\phi}\sqrt{w(\phi)}$, so that the eigenvalue equation takes the form $\mathcal{L}_{\text{SL},\phi}u_n(\phi) = -\lambda_n w(\phi) u_n(\phi)$; see, e.g., Appendix~B of \cite{Tomberg:2025fku}. However, the $\saL{\phi}$ operator makes the connection to the familiar wave function formalism of quantum mechanics more clear, and turns out to be useful in the computations below. The notation used here -- in particular, the minus sign in front of $\lambda$ and the definition of $w(\phi)$ -- hails from the Sturm--Liouville literature.} \cite{arfken2011mathematical, sturmliouville}. The eigenfunctions must satisfy the boundary conditions inherited from $P(\phi, N)$. For simple setups, they can be solved analytically; alternatively, they can be found numerically using a shooting method, distinguishing the different $n$ values by the number of zero-crossings. If the boundary conditions are such that there is no flux into the interval $[\phi_\text{low}, \phi_\text{high}]$, all eigenvalues are non-negative.

\paragraph{\boldmath $P$ decomposition.} Let us put the eigensolutions to use. Due to the completeness of the $u_n(\phi)$ basis, we can always write $P(\phi,N) = \sum_n a_n u_n(\phi) e^{-\lambda_n N}$, where the $a_n$ coefficients are set by the initial conditions, $a_n = \int \dd \phi \, \bar{u}_n(\phi) P(\phi, 0)$, by equation \eqref{eq:orthogonality}. In particular, if we fix $P(\phi, 0) = \delta(\phi-\phi_0)$, this gives
\begin{equation} \label{eq:P_decomposition}
\setlength{\fboxsep}{3\fboxsep}\boxed{
    P(\phi, N \cond \phi_0) = \sum_{n=1}^{\infty} \bar{u}_n(\phi_0)u_n(\phi)e^{-\lambda_n N} \, .
}
\end{equation}
\vspace{0.2cm}

\paragraph{\boldmath $\PFPT$ decomposition.}
Let us make one of the boundaries absorbing, denoted by $\phi_a$ as usual. Then, using \eqref{eq:PFPT} and \eqref{eq:P_decomposition}, we can decompose the first-time-passage distribution as
\begin{equation} \label{eq:PFPT_decomposition}
\setlength{\fboxsep}{3\fboxsep}\boxed{
    \PFPT(N, \phi_0)
    = \pm\sum_{n=1}^{\infty} \bar{u}_n(\phi_0) j_n(\phi_a) e^{-\lambda_n N} \, , 
}
\end{equation}
where $j_n(\phi)\equiv-\partial_\phi \qty[\frac{1}{2}\sigma^2(\phi)u_n(\phi)] + \mu(\phi)u_n(\phi)$ is the probability current component from $u_n(\phi)$ as given by \eqref{eq:j} (at the absorbing boundary, $j_n(\phi_a) = -\frac{1}{2}\sigma^2(\phi_a)\partial_{\phi_a}u_n(\phi_a)$), and the sign is inherited from \eqref{eq:PFPT}. The expected first-passage time becomes
\begin{equation} \label{eq:expN}
    \expval{N}_{\phi_0} \equiv \int_0^\infty \dd N \, N \PFPT(N, \phi_0)
    = \pm\sum_{n=1}^{\infty} \frac{1}{\lambda_n^2}\bar{u}_n(\phi_0) j_n(\phi_a) \, .
\end{equation}

We see that solving the eigenvalues and functions immediately yields useful results. Moreover, in the $N\to\infty$ limit, the lowest eigenvalue comes to dominate over all others in \eqref{eq:P_decomposition}--\eqref{eq:expN}, giving us easy access to the system's late-time behavior. The method is standard for solving linear partial differential equations but, perhaps surprisingly, hasn't been widely used to study enhanced inflaton perturbations and first-passage times in recent stochastic inflation literature (though applications to a spectator field distribution exist, see, e.g., \cite{Starobinsky:1994bd, Markkanen:2019kpv, Markkanen:2020bfc, Moreau:2020gib, Honda:2024evc, Nasuda:2025gyd}). In particular, result \eqref{eq:PFPT_decomposition} is, to the best of my knowledge, novel. Instead, recent work utilizes the \emph{characteristic function}, the Fourier transform of $\PFPT(N, \phi_0)$, which contains the same information. I compare the two methods in Section~\ref{sec:discussion}.

Next, I will apply these tools to the special case of constant-roll inflation in a hilltop potential.

\section{Constant roll around a hilltop}
\label{sec:cr}

\begin{figure}
    \centering
    \includegraphics{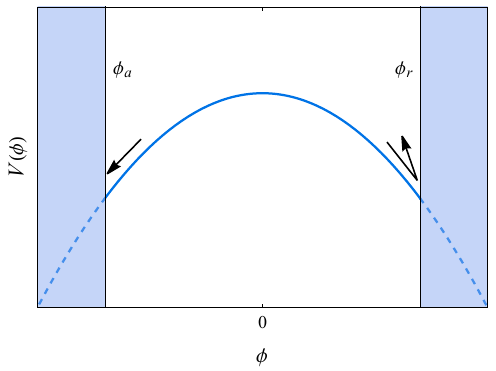}
    \caption{A sketch of the hilltop potential $V(\phi)$ from \eqref{eq:hilltop_eom}, with the absorbing and reflecting boundaries $\phi_a$ and $\phi_r$.}
    \label{fig:V_sketch}
\end{figure}

Let us consider inflation in a hilltop potential, with the equations of motion
\begin{equation} \label{eq:hilltop_eom}
    \ddot{\phi} + 3H\dot{\phi} + V'(\phi) = 0 \, , \quad 3H^2 = \frac{1}{2}\dot{\phi}^2 +  V(\phi) \, , \quad
    V(\phi) = V_0\qty(1+\frac{1}{2}\eta_V\phi^2) \, ,
\end{equation}
where dot denotes derivative with respect to the cosmic time $t$. The potential $V(\phi)$ has a maximum at $\phi=0$; it is sketched in Figure~\ref{fig:V_sketch} and characterized by two constants, the height $V_0$ and the second potential slow-roll parameter $\eta_V=V''(0)/V(0) < 0$ on the hilltop. We consider slow movement near the hilltop, so that the Hubble parameter $H\approx\sqrt{V(\phi)/3}$ can be approximated as constant (see, e.g., \cite{Karam:2022nym} for a similar setup). In terms of the number of e-folds $N$, with $\dd N = H \dd t$, equations \eqref{eq:hilltop_eom} then become
\begin{equation} \label{eq:hilltop_eom_N}
    \partial_N^2 \phi + 3\partial_N \phi + 3\eta_V\phi = 0 \, ,
\end{equation}
with the classical solution
\begin{equation} \label{eq:cr_solution}
    \phi(N) = c_+e^{A_+N} + c_-e^{A_-N} \xrightarrow{N \to \infty} c_+e^{A_+N} \, , \qquad A_\pm = \frac{3}{2}\qty(\pm\sqrt{1-\frac{4}{3}\eta_V} - 1) \, .
\end{equation}
For $\eta_V < 0$, we have $A_+ > 0$ and $A_- < 0$. In practice, the field converges fast onto an attractor solution, the `$+$' branch, where $\phi$ grows exponentially in $N$. On the attractor, the slow-roll parameters read
\begin{equation} \label{eq:cr_sr_parameters}
    \epsilon_1 \equiv \frac{\dot{\phi}^2}{2H^2} = \frac{1}{2}(\partial_N\phi)^2 = \frac{1}{2}A_+^2\phi^2 \, , \quad
    \epsilon_2 \equiv \partial_N \ln \epsilon_1 = 2A_+ \, .
\end{equation}
We see that $\epsilon_2$ is a (positive) constant: the system is in \emph{constant-roll} inflation.
With the full equations of motion, we would have $\epsilon_1 = -\partial_N \ln H$, so the constancy of $H$ is equivalent to $\epsilon_1 \ll 1$. This sets the restriction $|\phi| \ll 1/A_+$ for the field values under consideration, which is essentially equivalent to $\frac{1}{2}|\eta_V|\phi^2 \ll 1$, the requirement for the approximate constancy of the potential $V$ from \eqref{eq:hilltop_eom}. These restrictions don't play a significant role in the applications below.

Summing up, near a hilltop, the field follows the attractor behavior
\begin{equation} \label{eq:cr_attractor}
    \phi_\text{cl}(N) = \phi_0 e^{\frac{\epsilon_2}{2}N} \, , \qquad \partial_N \phi = \sqrt{2\epsilon_1} = \frac{\epsilon_2}{2}\phi \, ,
\end{equation}
where we introduced the notation `cl' to refer to the classical field motion, starting from $\phi_0$ at $N=0$.

\subsection{Stochastic evolution}
\label{sec:cr_stoch}
The above description is fully classical, but it serves as the basis for field evolution in stochastic inflation. In the presence of an attractor solution, stochastic motion happens along the attractor \cite{Tomberg:2022mkt}. This is true in slow-roll inflation, where $\epsilon_2 \ll 1$, but also in our constant-roll setup where $\epsilon_2$ can be large. The drift in the stochastic equation \eqref{eq:Langevin} arises from the classical motion \eqref{eq:cr_attractor},\footnote{In \cite{Tomberg:2025fku}, the same drift was written as $\mu(\phi)=-\beta\phi$, and both negative and positive $\beta$ were considered; here $b=-\beta$, and I only consider $b>0$ ($\beta<0$.)}
\begin{equation} \label{eq:cr_mu}
    \mu(\phi) = b\phi \, , \qquad b \equiv \frac{\epsilon_2}{2} = A_+ = \text{const.} > 0 \, ,
\end{equation}
while the diffusion coefficient arises from the quantum field perturbations. In constant roll, we have (see, e.g, Appendix E of \cite{Tomberg:2025fku}):
\begin{equation} \label{eq:cr_sigma}
    \sigma^2 = \frac{k^3_{\sigma_c}}{2\pi^2} |\delta\phi_{k_{\sigma_c}}|^2
    \overset{\text{CR}}{=}
    \frac{H^2}{4\pi^2}\frac{\sigma_c^3\pi}{2}|H_\nu(\sigma_c)|^2 = \text{const.} \, , \qquad \nu \equiv \frac{3}{2} + \frac{\epsilon_2}{2} \, .
\end{equation}
Here the constant $\sigma_c \ll 1$ sets the coarse-graining scale, so that a field Fourier mode $\delta\phi_k$ contributes its stochastic kick when $k=\sigma_c aH\equiv k_{\sigma_c}$ ($a$ is the scale factor); $\delta\phi_{k_{\sigma_c}}$ denotes the mode at coarse-graining, and I have evaluated it in the spatially flat gauge. $H_\nu$ is the Hankel function of the first kind. Using $\sigma_c \ll 1$ in the slow-roll limit, $\epsilon_2=0$, we recover the familiar result $\sigma^2 = H^2/(4\pi^2)$.

Importantly, $\sigma^2$ is a constant in \eqref{eq:cr_sigma}; in constant roll, the field perturbations are not affected by the (stochastic) background, so there is no backreaction between the two \cite{Tomberg:2023kli, Tomberg:2025fku}. The potential describes a (tachyonic) free field; the equations of motion are linear, independent for the background and perturbations.

It is noteworthy that classically, the attractor trajectory $\phi_\text{cl}$ \eqref{eq:cr_attractor} can't cross zero. However, stochastic trajectories can do this due to the diffusion term. This is not a problem: the drift \eqref{eq:cr_mu} and the diffusion coefficient \eqref{eq:cr_sigma} apply on both sides of the hilltop, assuming, as we do, that the field quickly settles on a new attractor once it crosses sides.

\paragraph{Boundaries.}
Stochastic constant-roll inflation near a hilltop was studied before in \paperI. Contrary to that work, I will add boundaries for the field evolution: an absorbing one below the hilltop and a reflecting one above the hilltop,
\begin{equation} \label{eq:cr_boundaries}
    \phi_\text{low}=\phi_a < 0 \, , \qquad 
    \phi_\text{high}=\phi_r > 0 \, .
\end{equation}
As discussed below equation \eqref{eq:FP_boundaries}, these correspond to dominant classical drift outside the hilltop, driving the field towards lower values and the end of inflation. The reflecting boundary acts like a local potential minimum, and the field may temporarily get trapped there. I will return to the physical interpretation of the model in Section~\ref{sec:discussion}, where I compare it to primordial black hole models with an inflection point potential.

\subsection{Rescaling}
\label{sec:rescaling}

Our stochastic constant-roll model has four dimensionful parameters: $b$, $\sigma$, $\phi_a$, and $\phi_r$. I eliminate the first two by introducing scaled field and time variables:
\begin{equation} \label{eq:varphi_varN}
    \varphi \equiv \frac{\sqrt{b}\phi}{\sigma} \, , \qquad
    \varN \equiv bN \, .
\end{equation}
The rescaled Fokker--Planck equation \eqref{eq:FP} then becomes
\begin{equation} \label{eq:FP_scaled}
    \partial_{\varN} P(\varphi, \varN) = \LFP{\varphi}P(\varphi, \varN) \equiv \partial_\varphi \qty[\partial_\varphi \qty(\frac{1}{2}
    P(\varphi, \varN)) - \varphi P(\varphi, \varN) ] \, ,
\end{equation}
so that effectively $b=\sigma=1$.
Below, I will solve the spectral decomposition of $\LFP{\varphi}$ in terms of the rescaled eigenvalues, 
\begin{equation} \label{eq:varlam}
    \varlam_n \equiv \frac{\lambda_n}{b} \, .
\end{equation}
I also introduce the notation
\begin{equation} \label{eq:varw}
    \varw(\varphi) \equiv \exp(-\varphi^2)
\end{equation}
for the rescaled version of $w(\phi)=\sigma^2\exp(-b\phi^2/\sigma^2)$ from \eqref{eq:self_adjoint_L}\footnote{The normalization of $w$ is, a priori, free; here, I have fixed it by choosing $\phi=0$ as the lower boundary of integration in \eqref{eq:self_adjoint_L}.}.

In the rescaled problem, all parameter dependence is shifted to the rescaled boundaries,
\begin{equation} \label{eq:rescaled_boundaries}
    \varphi_a \equiv \frac{\sqrt{b}\phi_a}{\sigma} \, , \qquad
    \varphi_r \equiv \frac{\sqrt{b}\phi_r}{\sigma} \, .
\end{equation}
For both, there are two qualitatively different regimes: the wide limit, $|\varphi_a|,\varphi_r \gg 1$, in which diffusion is weak and the field can remain insensitive to the boundaries for a long time; and the narrow limit, $|\varphi_a|,\varphi_r \ll 1$, where diffusion quickly drives a significant portion of the trajectories to the boundary. Scanning over $\varphi_a$ and $\varphi_r$ lets us study the problem in full generality; by undoing the scaling, one can always obtain a solution ($\lambda_n$ values and the corresponding eigenfunctions) for specific $b$, $\sigma$, $\phi_a$, and $\phi_r$.

\subsection{Eigenvalue problem}
\label{sec:eig_functions_and_values}

The eigenvalue equations \eqref{eq:all_eigensolutions} for the scaled constant-roll system are
\begin{align}
\label{eq:CR_eigeneq}
    \frac{1}{2}u_n''(\varphi) - \varphi u'_n(\varphi) - u_n(\varphi) &= -\varlam_n u_n(\varphi) \, , \\
\label{eq:CR_adjoint_eigeneq}
    \frac{1}{2}\bar{u}_n''(\varphi) + \varphi \bar{u}'_n(\varphi) &= -\varlam_n \bar{u}_n(\varphi) \, , \\
\label{eq:CR_v_eigeneq}
    \frac{1}{2}v_n''(\varphi) - \frac{1}{2}(1+\varphi^2)v_n(\varphi) &= - \varlam_n v_n(\varphi) \, .
\end{align}
Per \eqref{eq:all_eigensolutions}, the eigenfunctions are related by $v_n(\varphi) = \sqrt{\varw(\varphi)}u_n(\varphi) = \bar{u}_n(\varphi)/\sqrt{\varw(\varphi)}$; due to the varying factors of $\varw(\varphi)$, $u_n(\varphi)$ tends to grow exponentially for large $|\varphi|$, $\bar{u}_n(\varphi)$ tends to decay, and $v_n(\varphi)$ behaves moderately. It is enough to solve one of them. I choose to solve $\bar{u}_n(\phi)$ from \eqref{eq:CR_adjoint_eigeneq}; the general solution is \cite{Tomberg:2025fku}
\begin{equation} \label{eq:CR_eigenfunctions}
\begin{gathered}
    \bar{u}_n(\varphi) = A_n \bar{u}_{A,\varlam_n}(\varphi) + B_n \bar{u}_{B,\varlam_n}(\varphi) \, , \\
    \bar{u}_{A,\varlam}(\varphi) \equiv \iFi\qty(\frac{\varlam}{2}; \frac{1}{2}; -\varphi^2) \, , \quad
    \bar{u}_{B,\varlam}(\varphi) \equiv \varphi \cdot \iFi\qty(\frac{1+\varlam}{2}; \frac{3}{2}; -\varphi^2) \, .
\end{gathered}
\end{equation}
Here $\iFi$ is the confluent hypergeometric function (see, e.g., \cite{arfken2011mathematical}), with the series representation
\begin{equation} \label{eq:1F1_def}
    \iFi(a,b,z) = \sum_{n=0}^{\infty} \frac{a^{(n)}z^n}{b^{(n)}n!} \, ,
    \quad \text{where} \quad
    c^{(0)} \equiv 1 \, , \
    c^{(n)} \equiv c(c+1)(c+2)\dots(c+n-1) \, .
\end{equation}
The $A$ branch is even in $\varphi$, while the $B$ branch is odd. 

\begin{figure}
    \centering
    \includegraphics{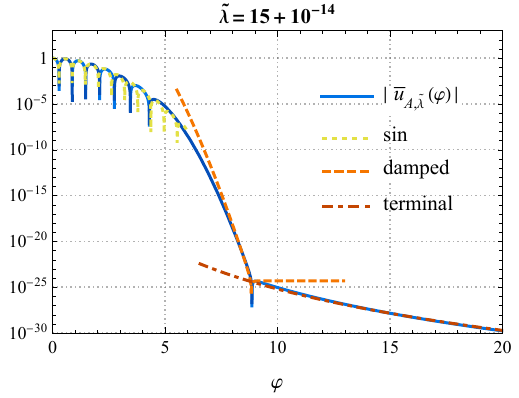}
    \caption{Behavior of the eigensolution $\bar{u}_{A,\varlam}$ for $\varlam = 15+10^{-14}$, compared to the various approximations from Section~\ref{sec:eig_functions}.}
    \label{fig:mode_behavior}
\end{figure}

An example solution is presented in Figure~\ref{fig:mode_behavior}. As expected, the solution oscillates in $\varphi$; $\varlam_n$ affects the oscillation frequency, with only discrete values satisfying the adjoint boundary conditions
\begin{equation} \label{eq:u_bar_boundaries}
    \bar{u}_n(\varphi_a) = 0 \, , \qquad
    \bar{u}'_n(\varphi_r) = 0 \, ,
\end{equation}
inherited from \eqref{eq:adj_FP_boundaries}. Plugging \eqref{eq:CR_eigenfunctions} in \eqref{eq:u_bar_boundaries} gives
\begin{equation} \label{eq:ubar_boundaries}
\begin{aligned}
    A_n \bar{u}_{A,\varlam_n}(\varphi_a) + B_n \bar{u}_{B,\varlam_n}(\varphi_a) &= 0 \, , \\
    A_n \bar{u}'_{A,\varlam_n}(\varphi_r) + B_n \bar{u}'_{B,\varlam_n}(\varphi_r) &= 0 \, ,
\end{aligned}
\end{equation}
and eliminating $A_n$ and $B_n$ yields
\begin{equation} \label{eq:lambda_equation}
    \bar{u}_{A,\varlam_n}(\varphi_a)\bar{u}'_{B,\varlam_n}(\varphi_r) - \bar{u}_{B,\varlam_n}(\varphi_a)\bar{u}'_{A,\varlam_n}(\varphi_r) = 0 \, .
\end{equation}
In practice, I find the $n$th eigenvalue $\varlam_n$ numerically as the $n$th solution of this oscillating equation. Plugging the eigenvalue back in \eqref{eq:ubar_boundaries} gives 
\begin{equation} \label{eq:A_B_ratio}
    \frac{A_n}{B_n} = -\frac{\bar{u}_{B,\varlam_n}(\varphi_a)}{\bar{u}_{A,\varlam_n}(\varphi_a)} \, ,
\end{equation}
which I combine with the normalization condition \eqref{eq:orthogonality},
\begin{equation} \label{eq:norms}
\begin{aligned}
    \int_{\varphi_a}^{\varphi_r} \frac{\bar{u}^2_n(\varphi)}{\varw(\varphi)} \dd \varphi
    &= A_n^2\int_{\varphi_a}^{\varphi_r} \frac{\bar{u}_{A,\varlam_n}^2(\varphi)}{\varw(\varphi)} \dd \varphi
    + B_n^2\int_{\varphi_a}^{\varphi_r} \frac{\bar{u}_{B,\varlam_n}^2(\varphi)}{\varw(\varphi)} \dd \varphi \\
    &\phantom{=} + 2 A_n B_n\int_{\varphi_a}^{\varphi_r} \frac{\bar{u}_{A,\varlam_n}(\varphi)\bar{u}_{B,\varlam_n}(\varphi)}{\varw(\varphi)} \dd \varphi = 1 \, ,
\end{aligned}
\end{equation}
to solve for $A_n$ and $B_n$. This procedure fully fixes $\varlam_n$ and $\bar{u}_n(\varphi)$ for each $n$.

\begin{figure}
    \centering
    \includegraphics{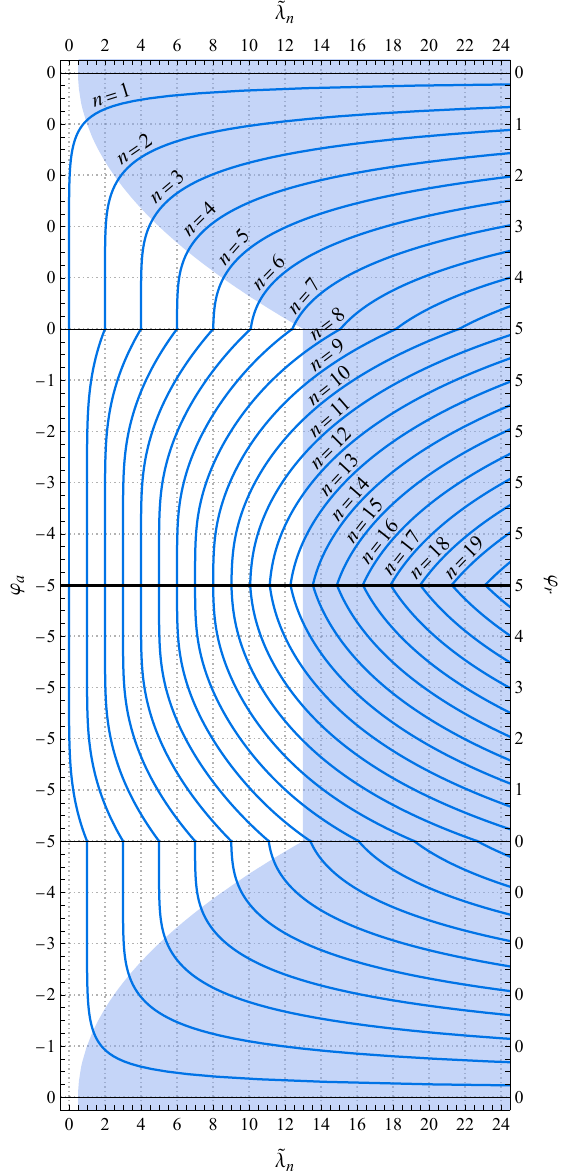}
    \caption{Lowest eigenvalues $\varlam_n$ (x-axis) in constant roll for varying boundaries $\varphi_a$ (left y-axis) and $\varphi_r$ (right y-axis). Moving from top to bottom, the varying boundaries draw a rectangle in the $(\varphi_a,\varphi_r)$ plane through the points $(0,0)$ (narrow limit), $(0,5)$ (half-wide limit), $(-5,5)$ (wide limit), $(-5,0)$ (half-wide limit), and back to $(0,0)$. The shaded region corresponds to $2\varlam_n - 1 > \qty(\max\{|\phi_a|,|\phi_r|\})^2$ (see the discussion around equation~\eqref{eq:lambda_sin}).}
    \label{fig:moving_boundaries}
\end{figure}

In Figure~\ref{fig:moving_boundaries}, I plot $\varlam_n$ for various values of the boundaries $\varphi_a$ and $\varphi_r$. Clear patterns emerge in the narrow and wide limits discussed below \eqref{eq:rescaled_boundaries}. In the rest of Section~\ref{sec:cr}, I explain these patterns by studying the behavior of the eigenfunctions in detail.

\subsection{Eigenfunction behavior}
\label{sec:eig_functions}

Let us return to Figure~\ref{fig:mode_behavior}, depicting an example solution of \eqref{eq:CR_adjoint_eigeneq} (the $\bar{u}_{A,\varlam}$ branch\footnote{In this section, I don't fix the boundaries and thus drop the index $n$ from the eigenfunctions and eigenvalues; the same solutions correspond to many different $n$ values, depending on the placement of the boundaries.}; the $B$ branch behaves similarly). We can distinguish three regimes in $\varphi$.

\paragraph{Sinusoidally oscillating regime.} At small $\varphi$, the solution oscillates. This is most easily understood in terms of $v$: when $\varphi^2 \ll 2\varlam-1$, \eqref{eq:CR_v_eigeneq} becomes
\begin{equation} \label{eq:oscillating_v}
    v''(\varphi) + \qty(2\varlam-1)v(\varphi) \approx 0 \quad \implies \quad
    v(\varphi) \approx c_1\sin\qty[\sqrt{2\varlam - 1}(\varphi - c_2)] 
\end{equation}
for some amplitude and phase $c_1$ and $c_2$. The same oscillations are inherited by $u$ and $\bar{u}$. Figure~\ref{fig:mode_behavior} shows that approximation \eqref{eq:oscillating_v}, scaled with $\sqrt{\varw(\varphi)}$, reproduces $\bar{u}(\varphi)$ well, once $c_1$ and $c_2$ have been fixed properly.

\paragraph{Damped regime.}
For $\varphi^2 \sim 2\varlam - 1$, the eigensolutions continue to oscillate for a while, but the oscillation frequency decreases, as can be seen in Figure~\ref{fig:mode_behavior}. In the $\bar{u}$ equation \eqref{eq:CR_adjoint_eigeneq}, the damping term $\varphi\bar{u}'(\varphi)$ grows in prominence. Once $\varphi^2 \ll 2\varlam$, the $\varlam$ term can be neglected altogether; the $\bar{u}$ equation becomes fully damped, 
\begin{equation} \label{eq:damped}
    \frac{1}{2}\bar{u}''(\varphi) +  \varphi \bar{u}'(\varphi) \approx 0 \quad
    \implies
    \quad
    \bar{u}'(\varphi) \approx \bar{c}_1 e^{-\varphi^2} \, , \quad \bar{u}(\varphi) \approx \bar{c}_2 + \bar{c}_3 \erf\varphi 
\end{equation}
for some integration constants $\bar{c}_1$, $\bar{c}_2$, and $\bar{c}_3$\footnote{Numerically, if $\bar{u}$ reaches very small values, it may be more convenient to work with the naturally small $\text{erfc} \, \varphi \equiv 1-\erf \varphi$ than $\erf\varphi$ (for large positive $\varphi$).}. For positive (negative) $\varphi$, if $\bar{c}_2$ and $\bar{c}_3$ have different (same) signs and $|c_3| > |c_2|$, the function crosses zero one more time in this region before approaching a constant ($\erf \varphi \xrightarrow{\varphi \gg 1} 1$). Figure~\ref{fig:mode_behavior} shows that \eqref{eq:damped} is a good approximation before and around the zero-crossing, when $\bar{c}_2$ and $\bar{c}_3$ are chosen properly.

\paragraph{Terminal regime.} According to \eqref{eq:damped}, 
 $\bar{u}(\varphi)$ approaches a constant for large $\varphi$, so that $\bar{u}''$ and $\bar{u}'$ decrease. Since the derivative terms in \eqref{eq:CR_adjoint_eigeneq} decay, the $\varlam$ term becomes important again, as long as the asymptotic $\bar{u}$ is non-zero. In practice, the eigenfunction settles to a `terminal velocity' regime, where the friction term balances the driving $\varlam$ term:
\begin{equation} \label{eq:terminal_solution}
    \varphi \bar{u}'(\varphi) \approx - \varlam \bar{u}(\varphi) \quad
    \implies
    \quad
    \bar{u}(\varphi) \approx \bar{c}_4\varphi^{-\varlam}
\end{equation}
for some $\bar{c}_4$.
Once $\bar{u}$ has settled into this regime, it can't get out. Solution \eqref{eq:terminal_solution} has no zero-crossings; it applies to all field values beyond the last oscillation, as we see in Figure~\ref{fig:mode_behavior}.

\paragraph{\boldmath Asymptotic behavior of $u_{A,\varlam}(\varphi)$ and $u_{B,\varlam}(\varphi)$.}
\label{sec:zero_crossings}
Let us connect the above discussion to the full solutions \eqref{eq:CR_eigenfunctions} for $\bar{u}_{A,\varlam}$ and $\bar{u}_{B,\varlam}$. Instead of $\bar{u}$, it is more convenient to work with the $u$ functions; using \emph{Kummer's transformation}, we obtain
\begin{equation} \label{eq:CR_eigenfunctions_2}
\begin{aligned}
    u_{A,\varlam}(\varphi)
    &= \bar{u}_{A,\varlam}(\varphi)/\varw(\varphi)
    = \iFi\qty(\frac{1-\varlam}{2}; \frac{1}{2}; \varphi^2) \, , \\
    u_{B,\varlam}(\varphi)
    &= \bar{u}_{B,\varlam}(\varphi)/\varw(\varphi)
    = \varphi \cdot \iFi\qty(1 - \frac{\varlam}{2}; \frac{3}{2}; \varphi^2) \, .
\end{aligned}
\end{equation}
In the series representation \eqref{eq:1F1_def} of these expressions, the $z^n$ and $b^{(n)}$ factors are positive, since $z=\varphi^2$ and $b=1/2$ ($A$) or $3/2$ ($B$) are positive, but $a^{(n)}$ can be either positive or negative, since $a=(1-\varlam)/2$ ($A$) or $1-\varlam/2$ ($B$) can be negative. In particular:
\begin{itemize}
    \item For $a>0$, all terms in the series are positive, and the series grows in an essentially exponential manner. The eigenfunction $u_{A,\varlam}(\varphi)$ doesn't cross zero, while $u_{B,\varlam}(\varphi)$ crosses zero exactly once at $\varphi=0$ due to the extra $\varphi$ factor.
    \item For $a<0$ but non-integer, the first $k$ factors in the rising factorial $a^{(n)}=a(a+1)(a+2)\dots(a+n-1)$ are negative, where $k$ equals the integer part of $|a|+1$, while the rest are positive. Terms with $n\leq k$ have alternating signs, translating into oscillations for small $\varphi$. Terms with $n > k$ have the fixed sign $(-1)^k$. The high-$n$ terms dominate the series for large $|\varphi|$, again leading to a growing (quasi-exponential) $|u(\varphi)|$, with the overall sign $(-1)^k$. The asymptotic sign flips when $a$ crosses non-positive integer values.
    
    \item At the `critical points,' $a \in \qty{0,-1,-2,\dots}$, the series has exactly $|a|+1$ non-zero terms, and then it terminates due to a zero in $a^{(n)}$, resulting in a polynomial, with the leading behavior $(-1)^{|a|}\varphi^{2|a|}$.
\end{itemize}
The critical points correspond to $\varlam = 1,3,5,\dots$ for the $A$ branch and $\varlam=2,4,6,\dots$ for the $B$ branch. At these points, new zero-crossings appear in $u_{A,\varlam}$ and $u_{B,\varlam}$ at $\varphi=\pm\infty$; as $\varlam$ increases, the zero-crossings move towards smaller $\varlam$ values. Correspondingly, at the critical points, the damped solution \eqref{eq:damped} vanishes at $\varphi=\pm\infty$, so there is no transition to the terminal regime; instead the damped region extends to infinity, with the polynomial $u(\varphi)$ from above (error-function-like $\bar{u}(\varphi)$ from \eqref{eq:damped}: for $\varphi>0$, $\bar{c}_2=-\bar{c}_3$ and $\bar{u}(\varphi)\sim \text{erfc}\,\varphi \xrightarrow{\varphi \to \infty}0$). All other $\varlam$ values exhibit a terminal regime, with the quasi-exponential $u(\varphi)$ from above (power-law $\bar{u}(\varphi)$ from \eqref{eq:terminal_solution}). Pushing $\varlam$ towards a critical value from above pushes the transition between the regimes towards larger $\varphi$ values. I chose $\varlam$ close to the critical value of $15$ in Figure~\ref{fig:mode_behavior} to make the damped region wide and easily distinguishable.

\subsection{Eigenvalues from boundary conditions}
\label{sec:eig_values}
Armed with an understanding of the eigenfunction behavior, we are now ready to present the analytical approximations for the eigenvalues $\varlam_n$ (and the related eigenfunctions) in various regimes of $\varphi_a$ and $\varphi_r$.

\begin{figure}
    \centering
    \includegraphics{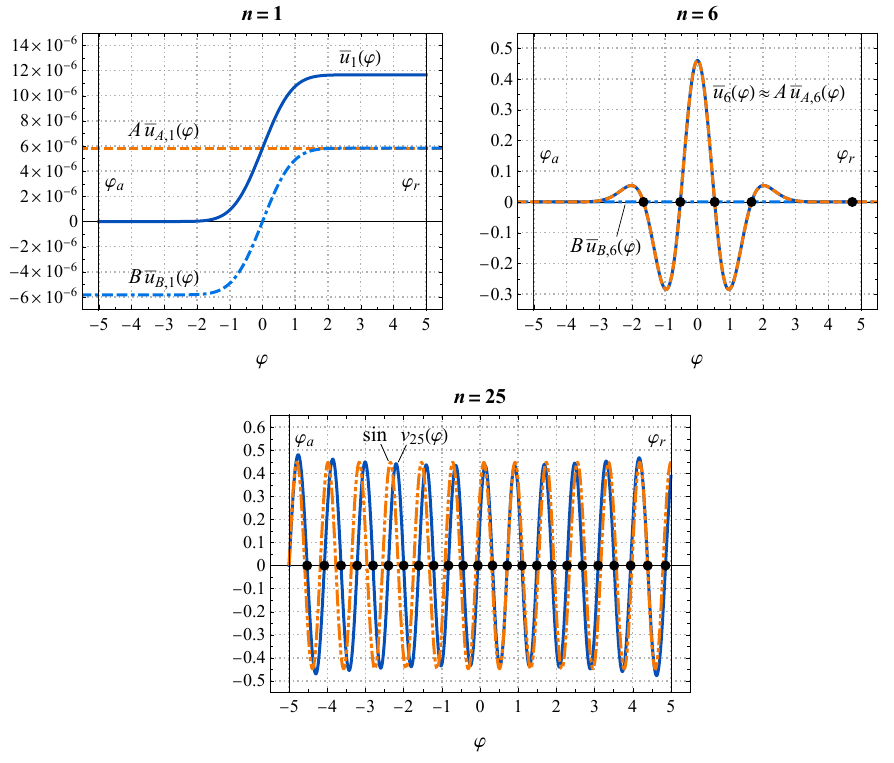}
    \caption{Example mode solutions for $n=1$, $n=6$ (in terms of $\bar{u}(\varphi)$ and its components), and $n=25$ (in terms of $v(\varphi)$) for the wide limits $\varphi_a=-5$, $\varphi_r=5$. The zero-crossings are marked with black dots (for $n=6$, the last zero-crossing is not clearly visible due to the exponential suppression of $\bar{u}$). For $n=1$, both components $A$ and $B$ contribute; for $n=6$, the $A$ branch dominates. For $n=25$, I compare $v(\varphi)$ with the sinusoidal approximation from \eqref{eq:oscillating_v}. For visual clarity, I have replaced $\varlam_n$ with $n$ in the subscripts.}
    \label{fig:modes}
\end{figure}

\subsubsection{Wide limit}
\label{sec:wide_limit}
The wide limit refers to $|\varphi_a|,\varphi_r \gg 1$\footnote{Note that, even in the wide limit, the field can still stay close enough to the hilltop for $H$ to be considered a constant. Below equation~\eqref{eq:cr_sr_parameters}, this requirement was shown to be equivalent to $\frac{1}{2}|\eta_V|\phi^2 \ll 1$. For $|\eta_V| \sim \mathcal{O}(1)$, and hence $b \sim \mathcal{O}(1)$ (see \eqref{eq:cr_mu} and \eqref{eq:cr_solution}), the requirement takes the rescaled form $|\varphi| \ll 1/\sigma \sim 1/H$. Since the Hubble parameter $H$ is vastly sub-Planckian in observationally allowed models, the requirement can be satisfied even for large $\varphi$ values.}. The behavior of the eigensolutions depends on $n$.

\paragraph{\boldmath $n=1$.}
\label{eq:lowest_eigenvalue}
According to Sturm--Liouville theory, the lowest eigenvalue is one with no zero-crossings in the interval $]\varphi_a, \varphi_r[$. To find it, let us note that, for $\varlam = 0$,
\begin{equation} \label{eq:lambda_0_solutions}
    \bar{u}_{A,0}(\varphi) = \iFi\qty(0;\frac{1}{2};-\varphi^2) = 1 \, , \qquad
    \bar{u}_{B,0}(\varphi) = \phi \cdot \iFi\qty(\frac{1}{2};\frac{3}{2};-\varphi^2) =  \frac{\sqrt{\pi}}{2}\erf(\varphi) \, .
\end{equation}
Since the error function obeys $\erf(x) \xrightarrow{x \lesssim -1} -1$, $\erf(x) \xrightarrow{x \gtrsim 1} 1$, we can construct the suitable solution without zero-crossings as
\begin{equation} \label{eq:lambda_0_u}
\setlength{\fboxsep}{5\fboxsep}\boxed{
    \bar{u}_{1}(\varphi) = A_1 \bar{u}_{A,0}(\varphi) + B_1 \bar{u}_{B,0}(\varphi) =
    C_1\qty(1 + \erf\varphi) = C_1\text{erfc} (-\varphi) \, ,
}
\end{equation}
which satisfies the boundary conditions \eqref{eq:u_bar_boundaries} for $\varphi_a \to -\infty$, $\varphi_r \to \infty$. This is, obviously, exactly of the damped form \eqref{eq:damped}, where we neglected the $\varlam$ term. For finite boundaries, this solution has to be perturbed slightly; nevertheless, \eqref{eq:lambda_0_u} remains a good approximation, up to the normalization $C_1=A_1=\frac{\sqrt{\pi}}{2}B_1$ to be determined from \eqref{eq:orthogonality}, with
\begin{equation} \label{eq:wide_lambda_1}
\setlength{\fboxsep}{5\fboxsep}\boxed{
    \varlam_1 \equiv \delta\varlam_1 \ll 1 \, .
}
\end{equation}
The leading eigenvalue is small: the field may become trapped in the depression near the reflecting boundary, taking a long time ($\sim 1/\varlam_1$) to tunnel out.

Figure~\ref{fig:modes} depicts the $n=1$ solution in an example case. Figure~\ref{fig:moving_boundaries} confirms that, in the wide limit, $\varlam_1$ is indeed close to zero.

\paragraph{\boldmath Intermediate $n$.}
\label{sec:wide_intermedtiate}
For $n\gtrsim 1$, the boundary conditions are satisfied for $\varlam$ close to, but slightly above, a critical value of one of the $A$/$B$ branches. There, the critical branch develops new zero-crossings at high $|\varphi|$, which is the correct behavior for the absorbing boundary, and `almost right' for the reflecting boundary -- indeed, $\bar{u}'(\varphi)$ vanishes at a local maximum of $\bar{u}(\varphi)$ close to the last zero-crossing, see Figure~\ref{fig:mode_behavior}. The critical branch dominates, while the other branch provides a subdominant contribution, adjusting the solution to match both boundary conditions exactly.\footnote{As a consistency check, note that according to the considerations of Section~\ref{sec:eig_functions}, the dominant branch has $n$ zero-crossings on the full real axis $\varphi \in ]-\infty, +\infty[$; one of these corresponds to the absorbing boundary at $\varphi_a$, and the rest land in $]\varphi_a, \varphi_r[$, giving the correct $n-1$ crossings for the eigenfunctions.}

From the dominant branch, we get
\begin{equation} \label{eq:wide_solutions}
\setlength{\fboxsep}{5\fboxsep}\boxed{
\begin{gathered}
    \varlam_{n} = n-1 + \delta\varlam_n \, ,
    \quad \delta\varlam_n \ll 1 \, , 
    \quad n=2,3,4,\dots \, , \\
    \bar{u}_{n}(\varphi) \approx
    \begin{cases}
        A_n \bar{u}_{A,\varlam_n}(\varphi) \, , & n=2,4,6,\dots \\
        B_n \bar{u}_{B,\varlam_n}(\varphi) \, , & n=3,5,7,\dots
    \end{cases}
\end{gathered}
}
\end{equation}
where $A_n$ and $B_n$ must be determined from the normalization condition \eqref{eq:orthogonality}. The eigenvalues are approximately equally spaced, like those of a quantum harmonic oscillator; indeed, equation \eqref{eq:CR_v_eigeneq} matches the time-independent Schrödinger equation of a harmonic oscillator when the boundaries are neglected (see, e.g., \cite{griffiths_introduction_2018}).

The small numbers $\delta\varlam_n$ in \eqref{eq:wide_lambda_1}  and \eqref{eq:wide_solutions} can be solved iteratively from \eqref{eq:lambda_equation}, e.g., using Newton's method. Due to their smallness, the iteration converges quickly. The iterative corrections can be expressed in terms of various generalized hypergeometric series of $\varphi_a$ and $\varphi_r$, but the explicit forms are not particularly informative. Instead, I list the numerical $\varlam_n$ and $\delta\varlam_n$ values in an example case in Appendix~\ref{sec:coefficient_list}, together with the normalization constants $A_n$ and $B_n$.

Figure~\ref{fig:modes} shows an example of an intermediate solution ($n=6$), verifying the $\bar{u}_n(\varphi)$ approximation of \eqref{eq:wide_solutions}. The near-integer $\varlam_n$ values of \eqref{eq:wide_solutions} are evident in the wide limit in Figure~\ref{fig:moving_boundaries}.

\paragraph{\boldmath Large $n$.}
\label{sec:high_lam_limit}
The above discussion applies if the boundaries $\varphi_a$, $\varphi_r$ are in the eigenfunctions' damped regime, near the outermost zero-crossings. According to the discussion of Section~\ref{sec:eig_functions}, this requires $\varphi_a^2,\varphi_r^2 \gtrsim 2\varlam - 1$. When $\varlam$ grows large, this condition breaks, even for wide boundaries. For large $\varlam_n$ and $n$, the full interval $[\varphi_a,\varphi_r]$ is in the sinusoidally oscillating regime, and the eigenfunctions follow \eqref{eq:oscillating_v}; the absorbing boundary condition fixes the phase $c_2=\varphi_a$, while the reflecting boundary, with $v'(\varphi_r) = \varphi v(\varphi_r)$ (equivalent to $\bar{u}'(\varphi_r)=0$), fixes the $\varlam_n$ spectrum:
\begin{equation} \label{eq:v_refl_boundary}
\begin{aligned}
    \cos(\sqrt{2\varlam_n - 1}(\varphi_r - \varphi_a))
    &= \frac{\varphi_r}{\sqrt{2\varlam_n - 1}}\sin(\sqrt{2\varlam_n - 1}(\varphi_r - \varphi_a)) \\
    &\approx 0 \ \text{for} \ 2\varlam_n - 1 \gg \varphi_r^2 \, 
\end{aligned}
\end{equation}
so that we get
\begin{equation} \label{eq:lambda_sin}
\setlength{\fboxsep}{5\fboxsep}\boxed{
\begin{gathered}    
    \varlam_n \approx \frac{\qty(n-\frac{1}{2})^2\pi^2}{2(\varphi_r-\varphi_a)^2} + \frac{1}{2} \, , \quad
    n \in \mathbb{Z}_+ \, , \quad
    n \gg \frac{1}{\pi}\max\{|\varphi_a|,|\varphi_r|\} (\varphi_r-\varphi_a) + \frac{1}{2}\, , \\
    v_n(\varphi) \approx \sqrt{\frac{2}{\varphi_r - \varphi_a}} \sin(\sqrt{2\varlam_n - 1} \, (\varphi - \varphi_a)) \, .
\end{gathered}
}
\end{equation}
The normalization of $v_n(\varphi)$ was set by \eqref{eq:orthogonality}, and the $n$ limit arises from $\varphi_a^2,\varphi_r^2 \ll 2\varlam_n - 1$. In this limit, contrary to the low $\varlam_n$ case, the system feels the effect of the boundaries, and $\varphi_r$ and $\varphi_a$ feature in the solutions. I chose $n$ so that $v_n(\varphi)$ has $n-1$ zero-crossings, in accordance with Sturm--Liouville theory.

\begin{figure}
    \centering
    \includegraphics{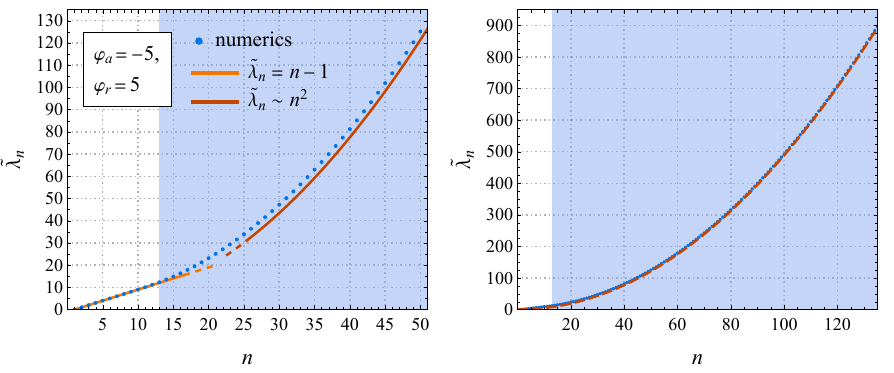}
    \caption{Eigenvalues $\varlam_n$ in the wide limit, $-\varphi_a=\varphi_r=5$. The points were obtained from a numerical solution of \eqref{eq:lambda_equation}; the lines depict the analytical approximations \eqref{eq:wide_solutions} ($\varlam_n=n-1$) and \eqref{eq:lambda_sin} ($\varlam_n\sim n^2$). The shaded region corresponds to $2\varlam_n - 1 > |\varphi_a|^2=|\varphi_r|^2$.}
    \label{fig:many_lambdas}
\end{figure}

Figure~\ref{fig:modes} depicts a sinusoidally oscillating example solution ($n=25$), showing a good match with the approximation \eqref{eq:lambda_sin}. In Figure~\ref{fig:moving_boundaries}, the region of applicability of \eqref{eq:lambda_sin} is shaded in blue; there, $\varlam_n$ clearly deviates from the integer behavior of \eqref{eq:wide_solutions}. Figure~\ref{fig:many_lambdas} plots many $\varlam_n$ values in an example model in the wide limit, showing good matches with both approximations \eqref{eq:wide_solutions} and \eqref{eq:lambda_sin} in their regions of applicability. The \eqref{eq:lambda_sin} approximation slightly underestimates the true eigenvalues for small $n$, but the relative error decreases as $n$ increases.

\subsubsection{Narrow limit}
\label{sec:narrow_limit}
In the narrow limit, where both $|\varphi_a|$ and $\varphi_r$ are small, and the validity condition of \eqref{eq:lambda_sin} applies for all $n$. This is reflected in the shaded region at the bottom and top of Figure~\ref{fig:moving_boundaries}. In the narrow limit, all eigensolutions follow the sinusoidal form \eqref{eq:lambda_sin}; $\varlam_n$ diverges for all $n$ as $\varphi_a,\varphi_r \to 0$.

\subsubsection{Half-wide limit}
\label{sec:half_wide_limit}
In the `half-wide' limit, either $\varphi_a$ or $\varphi_r$ tends to zero, while the other one stays large. We can use selected results from the previous sections to find the eigensolutions.

\paragraph{\boldmath Wide $\varphi_a$.}
For $\varphi_a \ll -1$, $\varphi_r \approx 0$, the even $A$ branch approximately satisfies the reflecting boundary condition $\bar{u}'_n(\varphi_r)=0$. It dominates all eigensolutions, with the $B$ branch providing subdominant corrections (when $\varphi_r \neq 0$) to exactly fix the boundary conditions. In other words,
\begin{equation} \label{eq:refl_dominant_solutions}
    \varlam_n = 2n-1 + \delta\varlam'_n \, ,
    \quad \delta\varlam'_n \ll 1 \, ,
    \quad \bar{u}_n(\varphi) \approx A\bar{u}_{A,\varlam_n}(\varphi) \, ,
    \quad n=1,2,3,\dots
\end{equation}
Only half of the wide limit's eigenvalues survive (those with an odd $\varlam_n$); in particular, the lowest eigenvalue now lies at $\varlam_1 \approx 1$ instead of $\varlam_1 \approx 0$. Since the reflecting boundary is at the top of the potential hill, the field won't get stuck in its vicinity for long.

\paragraph{\boldmath Wide $\varphi_r$.}
For $\varphi_a \approx 0$, $\varphi_r \gg 1$, the odd $B$ branch approximately satisfies the absorbing boundary condition $\bar{u}_n(\varphi_a)=0$. It dominates all eigensolutions, with the $A$ branch providing subdominant corrections to exactly fix the boundary conditions (when $\varphi_a \neq 0$). In other words,
\begin{equation} \label{eq:abs_dominant_solutions}
    \varlam_n = 2n-2 + \delta\varlam''_n \, ,
    \quad \delta\varlam''_n \ll 1 \, ,
    \quad \bar{u}_n(\varphi) \approx B\bar{u}_{B,\varlam_n}(\varphi) \, ,
    \quad n=1,2,3,\dots \, .
\end{equation}
Now the other half of the wide-limit solutions survive ($\varlam_n$ even); in particular, the lowest solution with $\varlam_1 \approx 0$ still exhibits the trapping behavior near the far-away reflecting boundary.

In both cases, the integer-$\varlam_n$ behavior only applies for small and intermediate $n$, while equation \eqref{eq:lambda_sin} describes the large-$n$ case. The transitions between the two regimes, as well as transitions between wide, half-wide, and narrow limits, can be seen in Figure~\ref{fig:moving_boundaries}.

\section{Time evolution of the field distribution}
\label{sec:time_evolution_of_P}

\begin{figure}
    \centering
    \includegraphics{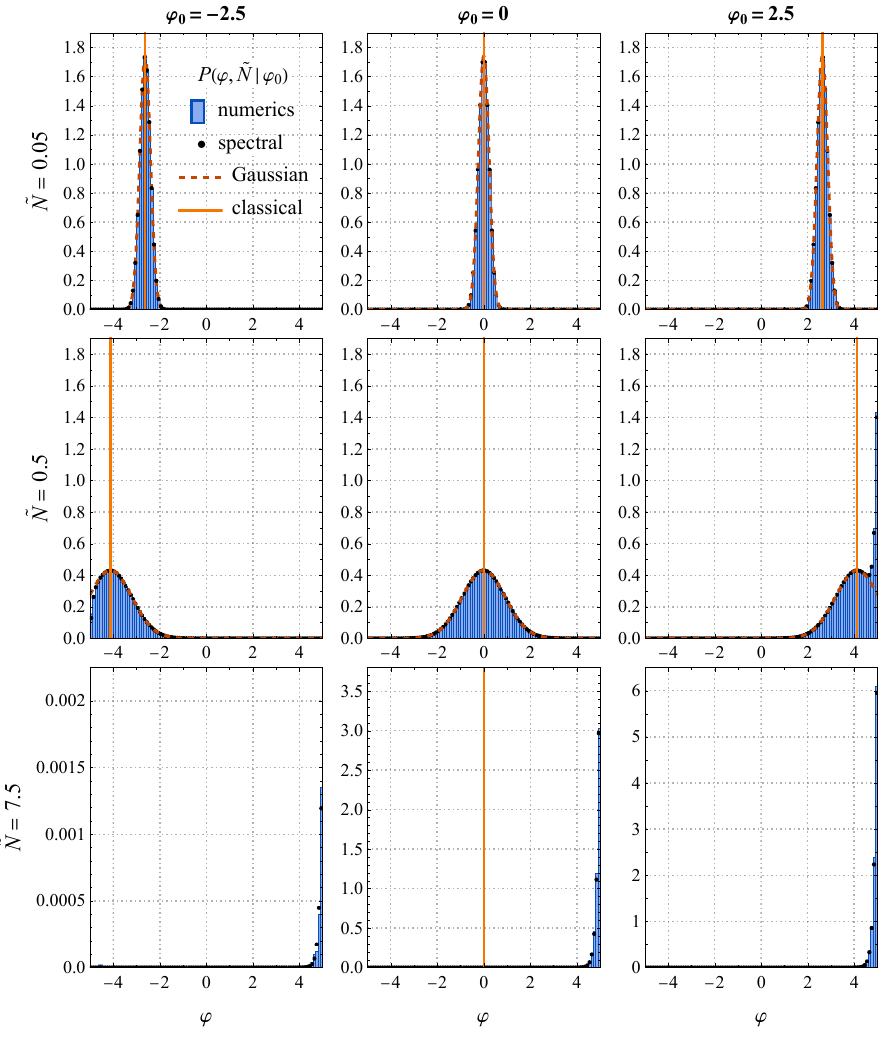}
    \caption{Probability distribution $P(\varphi,\varN\cond \varphi_0)$ in constant roll for various $\varphi_0$ and $\varN$, with absorbing and reflecting boundaries at $\varphi_a=-5$, $\varphi_r=5$. The blue histograms arise from numerically solving \eqref{eq:Langevin}; the black points are computed with the spectral decomposition \eqref{eq:P_decomposition}. The match is excellent for $\varN = 0.05$ and $0.5$; for $\varN=7.5$, there is a slight discrepancy due to the numerical results becoming inaccurate when only few points remain unabsorbed. The dashed and solid orange lines correspond to the Gaussian approximation \eqref{eq:P_Gauss} and the classical behavior \eqref{eq:cr_attractor}.}
    \label{fig:P_grid}
\end{figure}

To demonstrate the spectral method, I have solved the constant-roll system of Section~\ref{sec:cr} in an example case with wide boundaries, $\varphi_a = -5$, $\varphi_r = 5$. The large $|\varphi_a| = \sqrt{b}|\phi_a|/\sigma$ ensures that the evolution is drift-dominated at the absorbing boundary, $\mu(\phi_a) = b\phi_a \gg \sigma$ for $b \sim \mathcal{O}(1)$, so that stochasticity can be neglected between it and the end of inflation; meanwhile, the large $\varphi_r$ ensures that trajectories get stuck near the reflecting boundary for a long time, leading to interesting dynamics. I compare field distributions $P(\varphi, \varN \cond \varphi_0)$ obtained in three different ways:
 \begin{itemize}
     \item With the spectral decomposition \eqref{eq:P_decomposition}, using 142 modes (all modes with $\varlam_n \leq 1000$). The constants $\varlam_n$, $A_n$, and $B_n$, obtained from equations \eqref{eq:lambda_equation}--\eqref{eq:norms}, are listed in Appendix~\ref{sec:coefficient_list}. The equations were solved with Mathematica, which, in particular, has efficient implementations of the confluent hypergeometric functions. The analytical approximations of Section~\ref{sec:wide_limit} helped in finding the zero-crossings of the l.h.s. of \eqref{eq:lambda_equation}.
     \item By numerically solving the corresponding Langevin equation \eqref{eq:Langevin}\footnote{The rescaled Langevin equation reads $\frac{\dd \varphi}{\dd \varN} = \varphi + \xi(\varN)$.} $10^6$ times with random realizations of the noise, with the correct absorbing and reflecting boundaries and various initial conditions $\varphi_0$, and storing the field distribution at various time steps. The numerical solutions use a discrete time step of $10^{-3}$ in $\varN$ and a bin width $0.1$ for the stored $\varphi$ distributions. See, e.g., \cite{Tomberg:2024evi} for details on discretizing \eqref{eq:Langevin}.
     \item By ignoring the boundary conditions and solving the Fokker--Planck equation \eqref{eq:FP_scaled} analytically. The solution is Gaussian,
\begin{equation} \label{eq:P_Gauss}
\begin{gathered}
    P_G(\varphi, \varN \cond \varphi_0) =
    \frac{1}{\sqrt{2\pi}\Omega(\varN)}\exp{-\frac{[\varphi - \varphi_\text{cl}(\varN)]^2}{2\Omega^2(\varN)}} \, , \\
    \Omega^2(\varN) \equiv \frac{1}{2}\qty(e^{2\varN} - 1) \, , \qquad
    \varphi_\text{cl}(\varN) \equiv \varphi_0e^{\varN} \, .
\end{gathered}
\end{equation}
\end{itemize}
The results are depicted in Figure~\ref{fig:P_grid} for three values of $\varphi_0$, at three different timesteps: at $\varN=0.05$, when the distribution is still tightly peaked near the initial value; at $\varN=0.5$, when diffusion has widened the distribution and drift has driven it towards the edge; and at $\varN=7.5$, when most realizations have exited through the absorbing boundary. The spectral and numerical results agree, up to the numerical results' accuracy, demonstrating the viability of the spectral method. They also agree with the Gaussian result \eqref{eq:P_Gauss} for $\varN=0.05$ and $\varN=0.5$, except near the boundaries, which the Gaussian result doesn't account for. This is a property of the wide limit: the bulk of field evolution is unaffected by the boundaries until late times. At late times, most realizations have either exited through the absorbing boundary or clustered near the reflecting one, from where they only tunnel out slowly, as discussed in Section~\ref{sec:wide_limit}. This can be seen on the last row, $\varN = 7.5$. In Figure~\ref{fig:P_grid}, I have also marked the classical attractor \eqref{eq:cr_attractor}, matching the peak of the Gaussian approximation \eqref{eq:P_Gauss}. For $\varphi_0=\pm2.5$, $\varphi_\text{cl}$ crosses the left or right boundary at $\varN =\ln 2 \approx 0.7$; for $\varphi=0$, it stays at zero.

After verifying the spectral decomposition, let us study it in more detail at early and late times.

\begin{figure}
    \centering
    \includegraphics{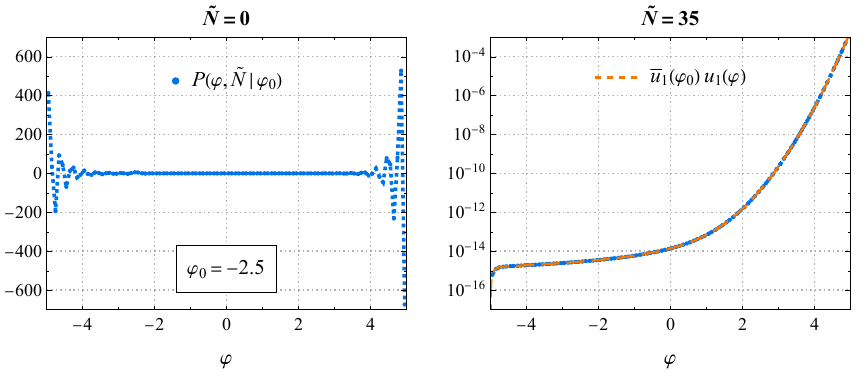}
    \caption{Probability distribution $P(\varphi,\varN\cond \varphi_0)$ in constant roll for $\varphi_0=-2.5$, with absorbing and reflecting boundaries at $\varphi_a=-5$, $\varphi_r=5$. The blue points are computed with the spectral decomposition \eqref{eq:P_decomposition}, up to $n=142$. The orange dashed line gives the first mode's contribution when $\varlam_1\varN_1 \ll 1$. At $\varN=0$, the spectral decomposition does not converge near the edges; at $\varN=35$, the leading mode dominates.}
    \label{fig:P_extreme_cases}
\end{figure}

\paragraph{Early times.} The left panel of Figure~\ref{fig:P_extreme_cases} depicts $P(\varphi, \varN \cond \varphi_0)$ computed using the spectral method at $\varN=0$, with $\varphi_0=-2.5$. The distribution should be a delta function centered around $\varphi=\varphi_0$; the distribution does have a bump at this value, but it is dominated by large oscillations near the boundaries. This is a convergence issue: at $\varN=0$, all eigenmodes are needed to formally produce the delta function \eqref{eq:completeness}, while our result only contains the first 142 modes.
The decomposition is similar in spirit to the well-known Fourier transform $\delta(x-x_0) = \int_{-\infty}^{\infty}\frac{\dd k}{2\pi}e^{ik(x-x_0)}$. The missing modes show up as uncanceled oscillations, which get amplified at large $|\varphi|$, since $u_n(\varphi)$ grows there, see Section~\ref{sec:eig_functions_and_values}.

For $\varN>0$, the modes decay as $e^{-\varlam_n \varN}$. Larger $n$ implies stronger suppression, and the spectral decomposition converges. A spectral decomposition truncated at $\varlam_\text{max}$ becomes accurate when
\begin{equation} \label{eq:N_cutoff}
    \varN \gg \frac{1}{\varlam_\text{max}} \, .
\end{equation}
With our $\varlam_\text{max}\approx 1000$, this becomes $\varN \gg 10^{-3}$. This is comfortably satisfied for $\varN = 0.05$ from Figure~\ref{fig:P_grid}, where the distribution is clearly well-converged.

\paragraph{Late times.} For large $\varN$, the large-$n$ modes become insignificant in the spectral decomposition \eqref{eq:P_decomposition}, one after another. Each mode is related to a time scale through $\varlam_n$, but also to a resolution scale in $\varphi$ through the oscillating eigenfunction $u_n(\varphi)$: as time goes on, the high-resolution modes decay away, leading to the widening of the distribution. In the end, only the $n=1$ mode survives, giving a universal, wide `steady state' distribution: the field is concentrated near the reflecting boundary $\varphi_r$, with a time-independent distribution shape $\bar{u}_1(\varphi_0)u_1(\varphi)$, decaying as $e^{-\varlam_1 \varN}$ as the field slowly leaks from $\varphi_r$ to $\varphi_a$. In our example case, the transition happens by $\varN \approx 30$; the right panel of Figure~\ref{fig:P_extreme_cases} verifies the behavior at $\varN = 35$. In the wide limit, $\varlam_1$ is extremely small, so $e^{-\varlam_1\varN} \approx 1$ at the time of transition.

\section{First-passage-time statistics}
\label{sec:FPT_statistics}

\begin{figure}
    \centering
    \includegraphics{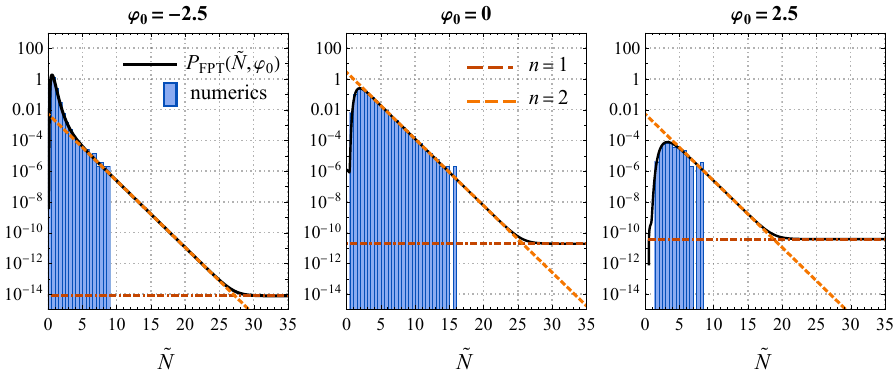}
    \caption{First-passage-time distribution $\PFPT(\varN,\varphi_0)$ in the setup of Figure~\ref{fig:P_grid}. The blue histograms arise from numerics; the black points are computed with the spectral decomposition \eqref{eq:PFPT_decomposition}. The match is excellent in the bulk region with enough statistics. The dashed lines depict the leading terms from \eqref{eq:PFPT_decomposition}, $-\bar{u}_n(\varphi_0)j_n(\varphi_a)e^{-\varlam_n \varN}$ for $n=1,2$.}
    \label{fig:PFPT_distributions}
\end{figure}

Let us next consider the distribution of first-passage times, $\PFPT(\varN,\varphi_0)$, through the absorbing boundary $\varphi_a$. I recorded the FPT times for the $10^6$ stochastic realizations discussed in Section~\ref{sec:time_evolution_of_P} and built their distribution for the three values of $\varphi_0$. Figure~\ref{fig:PFPT_distributions} shows the results, with $\varN$ bin width $0.05$, and compares them to the spectral decomposition \eqref{eq:PFPT_decomposition}. The match is again excellent in the bulk where $\PFPT(\varN,\varphi_0) \gtrsim 10^{-6}$.

The numerical results can't probe the large-$\varN$ tails of the distributions due to the limited statistics. On the other hand, the spectral decomposition only becomes more accurate for large $\varN$, as discussed above. This showcases the advantage of the spectral decomposition: it gives easy access to late-time statistics. For larger $\varN$, fewer terms contribute in the sum \eqref{eq:PFPT_decomposition}. In Figure~\ref{fig:PFPT_distributions}, I have plotted the first two contributions separately. After a few (rescaled) e-folds, the sum first becomes dominated by the $n=2$ term, and, around $\varN=20\dots 30$, the $n=1$ term starts to dominate. The mode amplitudes, and thus the transition times, depend on the initial field value through $\bar{u}_n(\varphi_0)$ in \eqref{eq:PFPT_decomposition}, but the qualitative behavior is always the same.

\subsection[Average \texorpdfstring{$N$}{N} and the \texorpdfstring{$\Delta N$}{Delta N} formalism]{\boldmath Average $N$ and the $\Delta N$ formalism}
\label{sec:delta_N}
The \emph{$\Delta N$ formalism} \cite{Sasaki:1995aw, Sasaki:1998ug, Wands:2000dp, Lyth:2004gb} relates the local amount of inflationary expansion to the local curvature perturbation $\zeta$:
\begin{equation} \label{eq:delta_N}
    \zeta = \Delta N \equiv N - N_\text{bg} \, ,
\end{equation}
where $N_\text{bg}$ is the expansion of the background. Schematically, the super-Hubble metric can be written as
\begin{equation}
    \dd s^2 = -\dd t^2 + a^2(t)e^{2\zeta(t,x)}\dd x^2
    = -\dd t^2 + e^{2N(t,x)}\dd x^2 \, ,
\end{equation}
where $a(t)=e^{N_\text{bg}(t)}$ is the background scale factor, defined to only depend on the cosmic time $t$ and not the spatial coordinates $x$; all spatial dependence is included in the curvature perturbation $\zeta(t,x)$, which is, essentially, a perturbation of $N_\text{bg}(t)$; the two combine to form the local total expansion $N(t,x)$.

In the \emph{stochastic $\Delta N$ formalism} \cite{Fujita:2013cna, Fujita:2014tja, Vennin:2015hra}, one starts with a Hubble-sized patch of unperturbed space with field value $\phi_0$ and evolves the field stochastically to the absorbing boundary $\phi_a$ at the end of inflation\footnote{The absorbing boundary can also be at a higher field value, at the edge of the quantum diffusion region; since there is no diffusion afterwards, the subsequent evolution proceeds at the same rate everywhere and doesn't contribute to $\Delta N$.}.
Different stochastic realizations describe Hubble-sized patches at different locations, branching from the original one as space expands. A patch's first-passage time through the boundary gives the $N$ in the $\Delta N$ formula \eqref{eq:delta_N}. Once we fix $N_\text{bg}$, \eqref{eq:delta_N} gives $\zeta$ for each patch.

Notably, the $\Delta N$ formalism itself does not determine $N_\text{bg}$, but leaves it as a matter of choice. If the perturbations are small, the choice is obvious: $N_\text{bg}$ should match the unperturbed evolution around which the perturbative expansion was made. In our constant-roll setup, this matches the classical number of e-folds on the attractor \eqref{eq:cr_attractor}:\footnote{For notational consistency, I will use the rescaled e-folds $\varN$ for the rest of Section~\ref{sec:FPT_statistics}. This does not change the qualitative discussion; $N$ and $\varN$ are related by \eqref{eq:varphi_varN}.}
\begin{equation} \label{eq:N_cl}
    \varN_\text{cl}(\varphi_0) \equiv \ln \frac{\varphi_a}{\varphi_0} \, .
\end{equation}
However, in the presence of large perturbations, defining a background is highly nontrivial. Let us explore different options.

\paragraph{Mean.}
In the stochastic $\Delta N$ formalism, $N_\text{bg}$ is customarily taken to be the stochastic mean of the individual patches' FPT times:\footnote{All averages presented in this section depend on the initial condition $\varphi_0$; for notational simplicity, I don't display it explicitly.}
\begin{equation} \label{eq:mean}
    \expvarN \equiv \int_0^\infty N \PFPT(\varN, \varphi_0)
    = -\sum_{n=1}^{\infty} \frac{1}{\varlam_n^2}\bar{u}_n(\varphi_0) j_n(\varphi_a) \, ,
\end{equation}
see \eqref{eq:expN}. With $\varN_\text{bg}=\expvarN$, we have $\expval{\zeta}=0$. However, if the spectrum contains very small eigenvalues, as is the case in the wide limit of our constant-roll system, the mean can grow large: the lowest eigenvalue dominates the sum in \eqref{eq:mean}, giving
\begin{equation} \label{eq:mean_approx}
    \expvarN \approx -\frac{1}{\varlam_1^2}\bar{u}_1(\varphi_0) j_1(\varphi_a) \equiv \expvarN_1  \, .
\end{equation}

\begin{figure}
    \centering
    \includegraphics{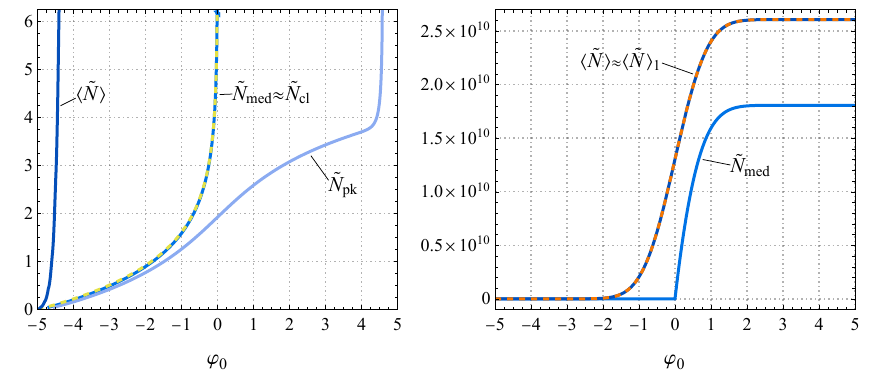}
    \caption{Average FPT statistics $\expvarN$, $\varN_\text{med}$, and $\varN_\text{pk}$ as functions of the initial field value $\varphi_0$ (solid lines), with the classical approximation $\varN_\text{cl}$ and the $\expvarN_1$ approximation from \eqref{eq:mean_approx} (dashed),
    in the constant-roll model with $\varphi_a=-5$, $\varphi_r=5$.}
    \label{fig:N_stas}
\end{figure}

The left panel of Figure~\ref{fig:N_stas} depicts $\expvarN$ as a function of $\varphi_0$ in our example model, confirming the validity of \eqref{eq:mean_approx}. The result is extremely large for all $\varphi_0$. For $\varphi_0 > 0$, this is expected: the field spends a long time stuck around the reflecting boundary before tunneling out. In fact, in this limit, the leading mode dominates from the beginning, and we can approximate $\PFPT(\varN,\varphi_0>0) \approx \varlam_1 e^{-\varlam_1 \varN}$, where the prefactor is set by the normalization of $\PFPT$. Comparing to the spectral decomposition \eqref{eq:PFPT_decomposition}, this implies\footnote{The first result in \eqref{eq:psotive_phi0_lemma} can also be derived directly by noting that
\begin{equation} \label{eq:lemma_derivation}
    -\bar{u}_1(\varphi_0)j_1(\varphi_a)
    = -\bar{u}_1(\varphi_0)\int_{\varphi_a}^{\varphi_r} \dd \varphi \, \LFP{\varphi}u_1(\varphi)
    = \varlam_1 \int_{\varphi_a}^{\varphi_r} \dd \varphi \, \bar{u}_1(\varphi_0) u_1(\varphi) \overset{\varphi_0>0}{\approx} \varlam_1 \int_{\varphi_a}^{\varphi_r} \dd \varphi \, \bar{u}_1(\varphi) u_1(\varphi) = \varlam_1 \, ,
\end{equation}
where I used $-\partial_\varphi j_n(\varphi) = \LFP{\varphi} u_n(\varphi) = -\varlam_n u_n(\varphi)$ (by definition), the fact that $u_1(\varphi)=\bar{u}_1(\varphi)e^{\varphi^2}$ only has significant support at large positive $\varphi$ values, where $\bar{u}_1$ is constant, so $\bar{u}_1(\varphi) \approx \bar{u}_1(\varphi_0)$, and the normalization condition \eqref{eq:orthogonality}.
}
\begin{equation} \label{eq:psotive_phi0_lemma}
    -\bar{u}_1(\varphi_0)j_1(\varphi_a) \approx \varlam_1 \, , \quad
    \expvarN \approx \frac{1}{\varlam_1} \quad \text{for} \ \varphi_0 > 0 \, .
\end{equation}
In our example model, $\varlam_1$ decreases and $\expvarN$ increases when $\varphi_r$ grows; for out chosen parameter values, $\expvarN \sim 10^{10}$ in this region. If the field interval was unbounded in the sense of $\phi_r \to \infty$, $\expvarN$ may even diverge; regulating such extreme cases through time-reversed stochastic inflation was recently discussed in \cite{Animali:2026omc, Blachier:2025iwk, Blachier:2025tcq}.

For $\varphi_0 > 0$, no classical trajectory can bring the field to the absorbing boundary.
In contrast, for $\varphi_0<0$, a classical trajectory exists; nevertheless, $\expvarN$ is much larger than the classical expectation $\varN_\text{cl}$  (for example, for $\varphi_0=-2.5$, $\expvarN\approx5.3\times10^6$ while $\varN_\text{cl}\approx0.7$). The bulk of $\varphi_0 < 0$ trajectories finish inflation around the classical time, but rare realizations venture to the other side of the hilltop and get stuck there. These rare patches skew the mean enormously due to their large FPT values.

Using $\expvarN$ as $\varN_\text{bg}$ in our example model is clearly absurd. If the observable universe started from a patch with $\varphi_0 < 0$, it is possible (and likely for some parameter values) that $\varphi$ never crosses the hilltop at any point inside the observable spatial region, yet the mere possibility of this happening is enough to dominate the sum \eqref{eq:mean}. Even if some points do cross the hilltop and get stuck on the other side, they can be so rare as to be inconsequential for describing the bulk of surrounding space. This behavior of $\expvarN$ is not surprising: it is well known that rare outliers can have a large impact on the mean, deviating it from the system's `typical' behavior. Let us explore other options for describing said behavior.

\paragraph{Median.} The median gives another way to measure the `average' of a distribution. It is the middle value of ordered samples, to be solved for the FPT from\footnote{Equivalently, we could write $\int_{0}^{\varN_\text{med}} \dd \varN \, \PFPT(\varN, \varphi_0) = \frac{1}{2}$; however, the form \eqref{eq:median} is more convenient for studying the late-time behavior, where only leading modes contribute, see \eqref{eq:median_approx}.}
\begin{equation} \label{eq:median}
    \int_{\varN_\text{med}}^{\infty} \dd \varN \, \PFPT(\varN, \varphi_0) =
    -\sum_{n=1}^{\infty} \frac{\bar{u}_n(\varphi_0) j_n(\varphi_a)}{\varlam_n} e^{-\varlam_n \varN}
    = \frac{1}{2} \, ,
\end{equation}
where I used \eqref{eq:PFPT_decomposition}. The median $\varN_\text{med}$ is less sensitive to rare outliers than the mean $\expvarN$.

The left panel of Figure~\ref{fig:N_stas} depicts $\varN_\text{med}$ as a function of $\varphi_0$ for $\varphi_0 < 0$ in our example model. In this regime, the median matches the classical behavior \eqref{eq:N_cl} extremely well. It is insensitive to the $\varphi > 0$ region of the potential; we would obtain essentially the same results if we brought the reflecting boundary down to $\varphi_r=0$.

In the wide limit, for $\varphi_0>0$, the median, too, will become dominated by the leading eigenmode, giving
\begin{equation} \label{eq:median_approx}
    \varN_\text{med} \approx \frac{1}{\varlam_1}\ln(-\frac{2\bar{u}_1(\varphi_0)j_1(\varphi_a)}{\varlam_1}) \approx \frac{\ln 2}{\varlam_1} \, ,
\end{equation}
where I used \eqref{eq:psotive_phi0_lemma}. This is similar to $\expvarN$ in the $\varphi_0>0$ region, of order $10^{10}$ in our example model and growing with $\varphi_r$, again with only weak $\varphi_0$ dependence: realizations with $\varphi_0>0$ end up in the vicinity of $\varphi_r$ fast, losing memory of the initial field value. I discuss techniques to refine approximation \eqref{eq:median_approx} in Appendix~\ref{sec:median_approx}; the full functional form of $\varN_\text{med}(\varphi_0)$ in the $\varphi_0>0$ region is shown in the right panel of Figure~\ref{fig:N_stas}.

In this work, I advocate using $\varN_\text{med}$ instead of $\expvarN$ for $\varN_\text{bg}$ in the $\Delta N$ formula. For $\varphi_0 < 0$, it better reproduces the `typical' behavior in the bulk of realizations, while still allowing, in principle, small corrections from stochastic effects. Since $\varN_\text{med}$ closely follows $\varN_\text{cl}$ in this regime, this method also matches linear perturbation theory for small perturbations, contrary to the version that uses $\expvarN$. For positive $\varphi_0$, the median grows large, signaling the fracturing of spacetime into patches with wildly different expansions. It is difficult to fix a meaningful background in such a complex spacetime; the median does as good a job as any other choice.

\paragraph{Mode.} For comparison, let us consider one more candidate for $\varN_\text{bg}$: the mode of the FPT distribution, that is, the value $\varN_\text{pk}$ where the distribution peaks, defined by
\begin{equation} \label{eq:mode}
    \partial_{\varN}\PFPT(\varN,\varphi_0)\big|_{\varN=\varN_\text{pk}} = 0 \, .
\end{equation}
In the left panel of Figure~\ref{fig:N_stas}, $\varN_\text{pk}$ is compared to $\varN_\text{med}$ and $\varN_\text{cl}$. The three agree for $\varphi_0$ values close to $\varphi_a$, but for larger $\varphi_0$, $\varN_\text{pk}$ is significantly lower. Since it deviates from $\varN_\text{cl}$, $\varN_\text{pk}$ is an unideal choice for $\varN_\text{bg}$. The peak location is also sensitive to the time variable used, since a change of variables can rescale the bin widths in ways that completely change the distribution shape; the integrated quantities $\expvarN$ and $\varN_\text{med}$ don't suffer from such a problem.

\bigskip

For narrow distributions, the different background conventions $\varN_\text{bg}=\expvarN, \varN_\text{med}, \varN_\text{pk}$ generally agree with each other. However, when the distribution develops heavy tails, like in our hilltop example, different conventions can produce wildly different results, as we saw above. I want to emphasize that the $\Delta N$ formalism itself does not set $N_\text{bg}$ -- it simply states that $\Delta N$, as defined in \eqref{eq:delta_N}, gives a reasonable non-linear generalization of the familiar curvature perturbation $\zeta$, with respect to some background. Fixing $N_\text{bg}$ amounts to fixing the background, a non-trivial task in a highly perturbed spacetime.

\subsection[Coarse-grained  \texorpdfstring{$\Delta N$}{Delta N}]{\boldmath Coarse-grained $\Delta N$}
\label{sec:coarse_grained_N}
In \paperI\ (see also \cite{Figueroa:2020jkf, Figueroa:2021zah, Tomberg:2024evi}), the $\Delta N$ formalism was used to compute the curvature perturbation at a fixed coarse-graining scale. In this approach, the stochastic evolution is followed until a fixed time $N_c$, and the subsequent evolution to $\phi_a$ is averaged over, yielding a time I denote by $N_\text{av}$. The coarse-grained curvature perturbation becomes
\begin{equation} \label{eq:coarse_grained_delta_N}
    \zeta_c = \Delta N_c = N_c + N_\text{av} - N_\text{bg} \, .
\end{equation}
The variation of $\zeta_c$ arises from the random value of $\phi$ at $N_c$, which I denote by $\phi_c$. Each $\phi_c$ value is related to a different averaged e-fold number $N_\text{av}$, making $N_\text{av}(\phi_c)$ in \eqref{eq:coarse_grained_delta_N} a stochastic variable. Only perturbation modes that have exited the coarse-graining scale by $N_c$ contribute to the variation of $\phi_c$, setting the final coarse-graining scale $k_c=\sigma_c a(N_c) H(N_c)$ (see discussion below equation \eqref{eq:cr_sigma}). 

Computation of $N_\text{av}(\phi_c)$, or its rescaled version $\varN_\text{av}(\varphi_c)$, faces the same ambiguity as the computation of $\varN_\text{bg}(\varphi_0)$ in Section~\ref{sec:delta_N}. It seems reasonable to use the same description for both. In \paperI, I used the classical number of e-folds, $\varN_\text{cl}$ from \eqref{eq:N_cl}. This restricted the allowed $\varphi_c$ to negative values, since $\varN_\text{cl}$ does not exist for the positive field values beyond the hilltop. Using one of the average quantities of Section~\ref{sec:delta_N}, we can now do better, extending the validity of \eqref{eq:coarse_grained_delta_N} to all $\varphi_c$ and hence all allowed $\Delta N$ values.

As advocated in Section~\ref{sec:delta_N}, I take $\varN_\text{bg}$ and $\varN_\text{av}$ to follow the median, \eqref{eq:median}. I take the initial field value to be negative, $\varphi_0 < 0$, so $\varN_\text{bg}(\varphi_0) \approx \varN_\text{cl}(\varphi_0)$. For $\varphi_c$, there are two main regimes of interest.

If $\varphi_c \lesssim 0$, too, then $\varN_\text{av}(\varphi_c) \approx \varN_\text{cl}(\varphi_c)$, and
\begin{equation} \label{eq:classical_delta_N_manipulation}
    \Delta \varN_c \approx \varN_c + \varN_\text{cl}(\varphi_c) - \varN_\text{cl}(\varphi_0)
    = \varN_c - \ln\frac{\varphi_c}{\varphi_0} \, .
\end{equation}
In the wide limit, we can use the Gaussian approximation \eqref{eq:P_Gauss} for the $\varphi_c$ distribution, yielding
\begin{equation} \label{eq:Gaussian_delta_N_distrib}
\begin{aligned}
    P_G(\Delta \varN_c)
    &= P_G(\varphi_c,\varN_c \cond \varphi_0) \qty|\frac{\dd \varphi_c}{\dd \Delta \varN_c}| \\
    &= \frac{1}{\sqrt{2\pi}\omega(\varN_c)}\exp{-\frac{\qty(1-e^{-\Delta \varN_c})^2}{2\omega^2(\varN_c)} - \Delta \varN_c} \, , \\
    \omega^2(\varN_c) &\equiv \frac{1}{2\varphi_0^2} \qty(1-e^{-2\varN_c}) \, ,
\end{aligned}
\end{equation}
where the subscript $G$ is inherited from \eqref{eq:P_Gauss}: \eqref{eq:Gaussian_delta_N_distrib} is a `modified Gaussian distribution,' related to a Gaussian one with a simple change of variables. This is the result derived in \paperI\footnote{To be more precise, \eqref{eq:Gaussian_delta_N_distrib} corresponds to equation (17) of \paperI, assuming the field perturbation power spectrum has frozen to a constant value $\sigma^2$, as it does in constant roll, see discussion around \eqref{eq:cr_sigma}. Then, the $\sigma_k^2$ factor in equation (17) of \paperI\ (not to be confused with the diffusion coefficient $\sigma$ of this paper, or the coarse-graining parameter $\sigma_c$) takes the form
\begin{equation} \label{eq:sigma_k_expression}
    \sigma_k^2 = \int_0^N \dd N' \, \PR(N')
    = \int_0^N \dd N' \, \frac{\sigma^2}{2\epsilon_1(N')}
    = \int_0^N \dd N' \, \frac{4\sigma^2}{\phi_0^2\epsilon_2^2}e^{-\epsilon_2 N'}
    =\frac{w^2(\varN)}{b^2}\, ,
\end{equation}
where $\PR(k)=\frac{k^3}{2\pi^2} \frac{|\delta\phi_k|^2}{2 \epsilon_1}=\frac{\sigma^2}{2\epsilon_1}$ is the curvature power spectrum, and I used \eqref{eq:cr_attractor}, \eqref{eq:cr_mu}, and \eqref{eq:varphi_varN} and switched variables between $N$ and $k$ (see \paperI\ for details). 
Using \eqref{eq:sigma_k_expression}, it is straightforward to verify that \eqref{eq:Gaussian_delta_N_distrib} and equation (17) of \paperI\ match. I will discuss a more complicated power spectrum with an initial transition phase in Section~\ref{sec:discussion}.
}. It exhibits a Gaussian center at small $\Delta \varN_c$, $P_G(\Delta \varN_c) \sim e^{-\Delta\varN_c^2/[2\omega^2(\varN_c)]}$, with an exponential tail, $P_G(\Delta\varN_c) \sim e^{-\Delta \varN_c}$ (with the rescaling removed, $P_G(\Delta N_c) \sim e^{-b \Delta N_c} = e^{-\frac{\epsilon_2}{2} \Delta N_c}$). Since $P_G(\Delta \varN_c)$ does not include the $\varphi_c > 0$ trajectories, \eqref{eq:Gaussian_delta_N_distrib} is not correctly normalized, though the error is typically small.

If $\varphi_c \gtrsim 1$, we have $\varN_\text{av} \approx \ln(2)/\varlam_1$ from \eqref{eq:median_approx}. This is a constant: $\varN_\text{av}$ is large (but finite) and almost independent of $\varphi_c$. Since $\varN_\text{av} \gg \varN_c, \varN_\text{bg}$ for typical $\varN_c, \varN_\text{bg} \sim \mathcal O(1)$, we have $\Delta \varN_c \approx \varN_\text{av}$ for all $\varphi_c > 0$: these trajectories form an approximate delta peak at $\Delta \varN_c = \Delta \varN_{c,\text{max}} \approx \ln(2)/\varlam_1$, the maximum allowed $\Delta \varN_c$ value. The existence of a maximal $\Delta \varN_c$ is due to the reflecting boundary at a finite field value $\varphi_r$; a similar maximum exists for all the averages from Section~\ref{sec:delta_N}\footnote{An upper limit in $\Delta N$ was earlier observed in hybrid inflation in \cite{Murata:2025onc}.}.

\begin{figure}
    \centering
    \includegraphics{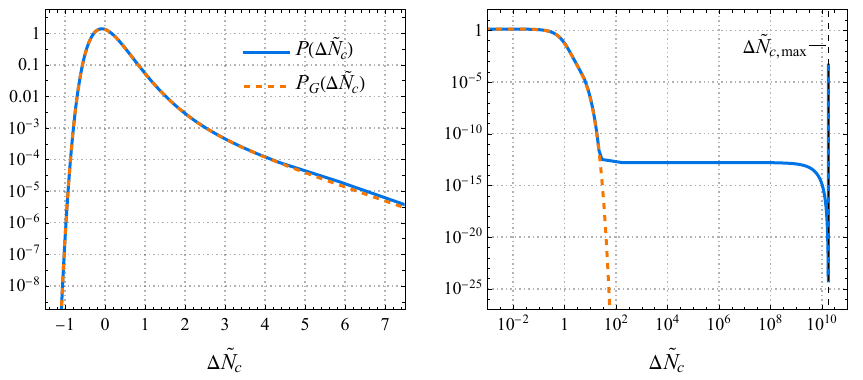}
    \caption{The probability distribution $P(\Delta \varN_c)$ in linear (left) and logarithmic (right) $\Delta \varN_c$, computed from the spectral decomposition using medians in the $\Delta N$ formalism, compared to the modified Gaussian approximation \eqref{eq:Gaussian_delta_N_distrib}. The parameter values are $\varphi_0=-1$, $\varN_c = 0.1$, $\varphi_a = -5$, and $\varphi_r = 5$.}
    \label{fig:Delta_N}
\end{figure}

Between the exponential tail and the delta peak, there is a transition region. The region starts at a small, negative $\varphi_c$, where $\varN_\text{av}$ diverges from $\varN_\text{cl}$. This can be estimated in various ways -- for example, by computing $\varN_\text{med}$ around $\varphi=0$ with the approximations of Appendix~\ref{sec:median_approx}, or by finding the $\varN$ where the first tunneling mode overtakes the `classical' ones in the $\PFPT$ distribution. In the transition region, $\varN_\text{med}$ grows fast over a narrow $\varphi$ range, see Figure~\ref{fig:N_stas}, leading to a plateau in $P(\Delta \varN_c)$ with the approximate height
\begin{equation} \label{eq:P_plateau_height}
    P_\text{trans} \approx P_G(\varphi=0,\varN_c\cond\varphi_0) \qty|\frac{\dd \varphi}{\dd \varN_\text{med}}|_{\varphi=0}
    \approx
    \frac{\varlam_1}{2\sqrt{2}
    \Omega(\varN_c)}\exp{-\frac{\varphi^2_\text{cl}(\varN_c)}{2\Omega^2(\varN_c)}} \, ,
\end{equation}
where I used the same logic as in equation \eqref{eq:Gaussian_delta_N_distrib}, but with $\varN_\text{med}$ evaluated using the approximations \eqref{eq:median_approx} and \eqref{eq:lambda_0_u}.

Figure~\ref{fig:Delta_N} depicts $P(\Delta \varN_c)$ in our example model with $\varphi_a=-5$, $\varphi_r=5$, $\varN_c = 0.1$, and $\varphi_0 = -1$, solved using the spectral decomposition. For small $\Delta \varN_c$, the distribution follows $P_G(\Delta \varN_c)$ from \eqref{eq:Gaussian_delta_N_distrib} extremely well, and for large $\Delta \varN_c$, we indeed see a second, delta-function-like peak.\footnote{A double-peak structure was earlier discovered in \cite{Kuroda:2025coa}, though there it arose from a curvaton mechanism.} The spectral decomposition is crucial for producing the distribution over the full $\Delta \varN_c$ range. The estimates above predict $\Delta \varN_{c,\text{max}} \approx 1.8\times 10^{10}$ and a transition region with $P_\text{trans} \approx 1.6\times 10^{-13}$ starting at $\Delta \varN_c \approx 20$; these describe the distribution well. We also see a dip in $P(\Delta \varN_c)$ just before $\Delta \varN_{c,\text{max}}$. This is caused by the declining $P(\varphi_c,\varN_c\cond\varphi_0)$ for positive $\varphi_c$. The trajectories accumulate near the reflecting boundary $\varphi_r$; there $P(\varphi_c,\varN_c\cond\varphi_0)$ grows again, deviating from the Gaussian form \eqref{eq:P_Gauss}. This accumulation enhances the peak at $\Delta \varN_{c,\text{max}}$.

\paperI\ suggested an upper cutoff $\Delta \varN_c \approx -\ln \omega(\varN_c)$ as the reliability limit of the modified Gaussian result \eqref{eq:Gaussian_delta_N_distrib}, based on the demand for $\varphi_c$ to be at least one standard deviation $\Omega(\varN_c)$ away from the problematic origin $\varphi=0$. In our example model, this limit becomes $\Delta \varN_c \approx 1.2$. However, we saw above that the median follows the classical evolution remarkably well even for field values very close to the origin, so that, with the median description, we can push the cutoff considerably further away, up to $\Delta \varN_c \approx 20$ in our example model.

\section{Discussion}
\label{sec:discussion}

Let me compare the above results to previous literature.

\paragraph{Primordial black hole models and comparison with \paperI.}
In \paperI, constant-roll inflation was studied in the context of primordial black hole models. Many such models feature an inflection point in the inflaton potential: a local minimum followed by a local maximum (see, e.g., \cite{Karam:2022nym} for a review). The inflaton first hits the minimum with high speed and then slows down as it climbs towards the maximum in an ultra-slow-roll-like phase. After crossing the hilltop, the field transitions to a dual constant-roll phase. The field's velocity is at its slowest just after the hilltop crossing, and this is also where the stochastic kicks are strongest, leading to an enhanced curvature perturbation. It is, in fact, enough to consider stochastic inflation in this region to obtain the curvature distribution at an excellent accuracy\footnote{This is true for a small coarse-graining parameter $\sigma_c$, see discussion around \eqref{eq:cr_sigma};
when $\sigma_c$ is pushed towards one, gradient effects become important, complicating the situation \cite{Jackson:2023obv, Raveendran:2025pnz}. In addition, stochasticity during the climb-up phase may become important for fine-tuned initial conditions \cite{Briaud:2023eae}.}. This is exactly what was done in \paperI, and again in the current work.

\paperI\ took the ultra-slow-roll-to-constant-roll transition into account when computing the stochastic noise. There, the noise coefficient $\sigma$ took the form
\begin{equation} \label{eq:old_noise}
    \sigma^2 = 2\epsilon_1(N)\PR(N,k_{\sigma_c}(N)) = \epsilon_2 \phi_0 e^{\frac{\epsilon_2}{2}N} \PR(N,k_{\sigma_c}(N)) \, ,
\end{equation}
where $\epsilon_1$, the first slow-roll parameter, is evaluated on the classical trajectory \eqref{eq:cr_attractor}; similarly, the curvature power spectrum $\PR$ is to be evaluated on the classical ultra-slow-roll-to-constant-roll trajectory. In practice, $\PR$ has frozen to its super-Hubble form by the time it contributes to the noise, so all scale dependence enters through the coarse-graining scale $k_{\sigma_c} = \sigma_c a H \propto e^{N}$ (see discussion around \eqref{eq:cr_sigma}). The spectrum peaks around the transition scale. After the transition, it drops as $\PR(k_{\sigma_c}) \propto k_{\sigma_c}^{-\frac{\epsilon_2}{2}} \propto e^{-\frac{\epsilon_2}{2}N}$ \cite{Karam:2022nym}, canceling with the prefactor in \eqref{eq:old_noise} and giving the constant $\sigma$ discussed in Section~\ref{sec:cr_stoch}. Before the transition, the spectrum rises approximately as $\PR \propto k^{4} \propto e^{4N}$ (with some dependence on prior evolution \cite{Byrnes:2018txb,Carrilho:2019oqg,Karam:2022nym}). In this regime, the noise is time-dependent. The corresponding Fokker-Planck equation \eqref{eq:FP} is no longer separable into $\phi$ and $N$ dependent parts, and the solutions don't take the simple form \eqref{eq:P_decomposition} (formally, $\bar{u}_n(\phi_0)$ are replaced by time-dependent coefficients). To use the spectral method in this case, one could solve \eqref{eq:FP} numerically until the peak scale has passed, and use the result as an initial condition for the subsequent, decomposable evolution with constant $\sigma$. In this paper, I omit such complications, taking $\sigma^2$ to switch from zero to the constant-roll expression \eqref{eq:cr_sigma} instantaneously at the field value $\phi_0$. This doesn't affect the system's qualitative long-time behavior.

\paperI\ only considered stochastic evolution on one slope of the hill (the side where inflation ends). This paper extends the analysis to trajectories that return to the other side of the hilltop. The key insight, discussed in Section~\ref{sec:cr_stoch}, is that stochastic evolution conforms
to a similar constant-roll attractor on both sides, as noted earlier in \cite{Tomberg:2025fku}. In particular, the field does \emph{not} return to the ultra-slow-roll behavior it previously had on the other side of the hill, since it has already lost the necessary kinetic energy. 

In \cite{Tomberg:2025fku}, absorbing boundaries were placed on both sides of the hill; here, I have made the upper boundary a reflecting one, mimicking the local potential minimum of a full PBH potential. The system's late-time behavior is dominated by the reflecting boundary. Obviously, the late-time behavior of a realistic PBH model differs somewhat from that of the reflecting constant-roll case, but the latter may provide valuable qualitative insights. In particular, rare patches of space may be expected to get stuck in the local minimum also in the realistic potential, taking a long time to tunnel out; such behaviour leads, in general, to problems with using the mean $\expval{N}$ as the background, as discussed in Section~\ref{sec:delta_N}. Using the median instead, one may then expect a local exponential tail in $\Delta N$ associated with a hilltop to be overtaken at even larger $\Delta N$ values by more complicated, peaked behavior associated with the minimum. The bounded hilltop model is the simplest realization of this behavior, perhaps the only such model that is solvable semi-analytically.

\paragraph{\boldmath Comparison with classical $\Delta N$.}
Instead of the stochastic formalism, some works evolve the field classically from perturbed initial conditions, computing the $\Delta N$ statistics from the classical trajectories, see, e.g., \cite{Cai:2018dkf, Atal:2019cdz, Atal:2019erb, Cai:2021zsp, Biagetti:2021eep, Hooshangi:2021ubn, Cai:2022erk, Hooshangi:2022lao, Pi:2022ysn, Firouzjahi:2023ahg, Escriva:2023uko, Escriva:2025ftp}. \paperI\ showed that, for constant roll near a hilltop, both methods produce the same modified Gaussian result \eqref{eq:Gaussian_delta_N_distrib} in the wide limit, when restricting attention to cases with $\phi_c, \phi_0 < 0$.

The relationship between the classical and stochastic $\Delta N$ formalisms was studied in detail in \cite{Ballesteros:2024pwn} for constant-roll models, using different approximations. I argued above that the field conforms to the constant-roll attractor on both sides of the hilltop, instead of returning to the ultra-slow-roll behavior. In the language of \cite{Ballesteros:2024pwn}, this corresponds to the \emph{eternal CR approximation} instead of the \emph{unperturbed trajectory approximation}.

As recognized in \cite{Ballesteros:2024pwn}, the classical $\Delta N$ approximation can't handle trajectories with $\phi_c>0$, since the classical evolution does not bring them to the end-of-inflation boundary (unless the field is far removed from the constant-roll attractor, with high negative initial field velocity).  This can be remedied in the stochastic formalism, as I did above by replacing the classical e-fold number with the median of the stochastic process. The large-perturbation, $\phi_c > 0$ regime is truly stochastic, ruled by quantum diffusion instead of the classical drift, and the results can't be reproduced by the random initial conditions of the classical $\Delta N$ formalism.

\paragraph{Comparison with the characteristic function formalism.}
Instead of the spectral method, recent work \cite{Pattison:2017mbe, Ezquiaga:2019ftu, Pattison:2021oen, Animali:2022otk, Animali:2024jiz} has used the \emph{characteristic function} method. The characteristic function $\chi$ is the Fourier transform of the FPT probability distribution,
\begin{equation} \label{eq:chi}
    \chi(\omega,\phi_0) \equiv \expval{e^{i\omega N}}_{\phi_0}
    = \int_0^\infty \dd N \, \PFPT(N,\phi_0) e^{i\omega N} \, .
\end{equation}
The adjoint Fokker--Planck equation \eqref{eq:PFPT_eq} can be written as
\begin{equation} \label{eq:chi_eom}
    \adLFP{\phi_0} \chi(\omega, \phi_0)
    = -i\omega \chi(\omega, \phi_0) \, ,
\end{equation}
an ordinary differential equation in $\phi$ for each value of the frequency variable $\omega$, with the boundary conditions $\chi(\omega, \phi_a) = 1$ (ensuring $\PFPT(N, \phi_a) = \delta(N)$\footnote{In \eqref{eq:PFPT_boundaries}, we used the boundary condition $\PFPT(N,\phi_a)=0$ for all $N$. To enforce the correct normalization $\int_0^\infty \dd N \, \PFPT(N,\phi_a) = 1$, we actually need to relax this for $N=0$, leading to the $\delta$-function boundary condition. The eigenfunctions still follow the boundary conditions $\bar{u}_n(\phi_a)=0$; the $\delta$-function arises non-trivially from the infinite sum \eqref{eq:PFPT_decomposition}.}) and $\delta_{\phi_r}\chi(\omega, \phi_r) = 0$. The solution takes the general form \cite{Ezquiaga:2019ftu}\footnote{In \cite{Ezquiaga:2019ftu}, an additional analytic piece $g(\omega,\phi_0)$ was included in $\chi(\omega,\phi_0)$; for the discrete spectrum arising from finite boundary conditions, it is not necessary.}
\begin{equation} \label{eq:chi_solution}
    \chi(\omega, \phi_0) = 
    \sum_n \frac{a_n(\phi_0)}{\lambda_n - i\omega} \, ,
\end{equation}
where $a_n(\phi_0)$ is an analytic function. Inverse Fourier transform gives $\PFPT$; contour integration transforms the poles at $\omega=-i\lambda_n$ into the decaying exponents of the spectral decomposition \eqref{eq:PFPT_decomposition} \cite{Ezquiaga:2019ftu}. From the residues, we can further identify $a_n(\phi_0) = \pm \bar{u}_n(\phi_0)j(\phi_a)$. The basics of this comparison were presented in \cite{Ezquiaga:2019ftu}; in this paper, I have developed the spectral method further, linking it also to the field distribution $P(\phi,N \cond \phi_0)$.

The characteristic function and spectral methods obviously produce the same results, but they have different strengths. The first relies on complex analysis, while the second uses real variables.  The characteristic function provides quick access to certain results, such as the expected first-passage time $\expval{N}_{\phi_0} = -i\partial_\omega \chi(\omega,\phi_0)\big|_{\omega=0}$ (obvious from \eqref{eq:chi}), especially in analytically solvable models. The spectral method first requires the solution of the eigenvalues and functions, which can be tedious -- though it is, at least in principle, straightforward to implement numerically using a shooting method. Once the eigensolutions have been obtained, the spectral method offers remarkable flexibility in playing with the initial condition $\phi_0$ and the time $N$ through the expansions \eqref{eq:P_decomposition}, \eqref{eq:PFPT_decomposition}.

\paragraph{Comparison with recent work on the spectral method.} When this paper was in the final stages of preparation, reference \cite{Mishra:2026xal} appeared, presenting the spectral method in stochastic inflation in detail. The results of \cite{Mishra:2026xal} are compatible with and complementary to those of the current paper. In \cite{Mishra:2026xal}, the authors derived an alternative expression for the spectral decomposition of $\PFPT(N,\phi_0)$; I leave the comparison between it and this paper's result \eqref{eq:PFPT_decomposition} for future work. In this paper, I extended the spectral method from $\PFPT(N,\phi_0)$ to the field distribution $P(\phi,N)$. Reference \cite{Mishra:2026xal} showcased the spectral method in a setup with constant (zero or non-zero) drift $\mu$, leaving the constant-roll case of $\mu(\phi) \propto \phi$ for future work; this is exactly the case studied in the current paper.

\section{Conclusions}
\label{sec:conclusions}

In this paper, I solved stochastic constant-roll inflation in a hilltop potential, with an absorbing boundary on one side of the hilltop (representing the end of inflation) and a reflecting boundary on the other (representing a steep classical regime). The solution includes stochastic trajectories that cross the hilltop; it assumes the field quickly approaches the constant-roll attractor on either side. I presented in detail the spectral method, which gives easy access to the system's asymptotic behavior, including the late-time field and first-passage-time distributions and, in particular, the tail of the coarse-grained $\Delta N$ distribution.

At early times, a large collection of the spectral eigenmodes contributes to the statistics. The mean field follows the classical trajectory; diffusion gives the field distribution a non-zero width around this classical mean. At late times, the lowest eigenmode becomes dominant -- the corresponding stochastic trajectories cross the hilltop and get stuck in the local potential minimum at the reflecting boundary, tunneling out only slowly. This regime is diffusion dominated and inherently stochastic, not reproducible with semi-classical approximations.

Even when the long-lasting late-time trajectories are rare, they turn out to completely dominate the mean of the first-passage time. If the initial field value is close to the absorbing boundary, typical trajectories reach the boundary quickly, in a number of e-folds approximately given by the classical evolution. The mean FPT is drastically larger. Contrary to what is typically assumed in the literature, the mean doesn't describe the bulk behavior well and should not be used as the background e-fold value in the $\Delta N$ formalism. Contrary to the mean, the median is less sensitive to the rare extremes and, indeed, follows the classical behavior up to initial field values very close to the hilltop. Beyond the hilltop, the median still gives sensible first-passage times, generalizing the classical result to the non-classical region. The issue of a sensible background in the diffusion-dominated regime was also raised in \cite{Blachier:2025iwk}.

On the absorbing side of the hilltop, the $\Delta N$ distribution has the well-known form of a Gaussian peak with an exponential tail, computed stochastically in \paperI\ and reproducible with the classical $\Delta N$ formalism. When the computation is extended beyond the hilltop using the median for the background, the exponential tail gets cut off, first flattening into a plateau and then forming a sharp spike at a large $\Delta N$ value, corresponding to the tunneling solution. This is the main result of the current paper.

The hilltop potential studied here is a proxy for primordial black hole models. In realistic models, the reflecting boundary is replaced by a smooth minimum in the potential, followed by a steep slope. While the details differ, one may expect qualitatively similar behavior: beyond the hilltop, the local minimum traps the field, so that new, smaller eigenvalues come to dominate, and the $\Delta N$ distribution develops a far-away peak.

Above, I used the median FPT for the background $N$ in the diffusion-dominated regime. Such a description may be useful for other cases with large perturbations. For example, for a semi-infinite quantum well, the mean of $N$ diverges (see, e.g., \cite{Ando:2020fjm, Blachier:2025iwk, Blachier:2025tcq}); the median is always finite, and could be used as a `regularized' background with respect to which one can define the curvature perturbations. However, while sensible, this is only one possible background description. Indeed, the diffusion-dominated regime is extremely volatile, related to eternal inflation \cite{Tomberg:2025fku}, where space-time follows a complicated fractal structure \cite{Aryal:1987vn, Winitzki:2001np, Jain:2019gsq}. This structure is inherently non-perturbative, and the $\Delta N$ formalism is not well-suited to describe it, since it relies on a clean division between background and perturbations. New methods may be needed to capture the full details.

\acknowledgments
The author thanks Chiara Animali, Christophe Ringeval, and Baptiste Blachier for helpful discussions. This work was supported by the ``Fonds de la Recherche Scientifique'' (FNRS) under the IISN grant number 4.4517.08.

\appendix

\section{Approximating the median}
\label{sec:median_approx}
Solving the median \eqref{eq:median} numerically requires repeated computations of a complicated sum. This is tedious, particularly for $\varphi_0 \gtrsim 0$, where $\varN_\text{med}$ quickly grows from order one to $1/\varlam_1 \sim 10^{10}$. Fortunately, as $\varN$ starts to grow, only the first few terms contribute significantly to the sum. In fact, we can obtain a good approximation by keeping just the two leading terms, approximating the survival probability as
\begin{equation} \label{eq:median_approx_setup}
\begin{gathered}
    \int_{\varN'}^{\infty} \dd \varN' \, \PFPT(\varN', \varphi_0) \equiv S(\varN \cond \varphi_0)
    \approx S_1(\varphi_0)e^{-\varlam_1 \varN} + S_2(\varphi_0)e^{-\varlam_2 \varN} \, , \\
    S_n(\varphi_0) \equiv -\frac{\bar{u}_n(\varphi_0) j_n(\varphi_a)}{\varlam_n} \, .
\end{gathered}
\end{equation}

For not-too-large $\varN_\text{med}$, that is, $\varlam_1\varN_\text{med} \ll 1$, we can expand the first exponential function in \eqref{eq:median_approx_setup} and solve
\begin{equation} \label{eq:median_approx_sol_1}
\begin{gathered}
    S(\varN_\text{med}) \approx S_1(\phi_0)\qty[1-\varlam_1 \varN_\text{med}] + S_2(\varphi_0)e^{-\varlam_2 \varN_\text{med}} = \frac{1}{2} \\
    \implies \quad
    \varN_\text{med} \approx \frac{1}{\varlam_2}W_0\qty(\frac{S_2(\varphi_0)\varlam_2}{S_1(\varphi_0)\varlam_1}e^{\frac{\varlam_2}{\varlam_1}\qty(\frac{1}{2S_1(\varphi_0)}-1)})
    -\frac{1}{\varlam_1}\qty(\frac{1}{2S_1(\varphi_0)}-1) \, ,
\end{gathered}
\end{equation}
where $W_0$ is the principal branch of the product logarithm. This is a good approximation in the vicinity of $\varphi_0=0$, at the edge between the classical behavior $\varN_\text{med}\approx\varN_\text{cl}$ and the diffusion-dominated behavior beyond the hilltop.

In the diffusion regime with large positive $\varphi_0$, $\varlam_1\varN_\text{med}$ grows, and the first term becomes increasingly dominant compared to the second in \eqref{eq:median_approx_setup}. Let us define
\begin{equation} \label{eq:median_approx_sol_2a}
\begin{gathered}
    \varN_\text{med} = \varN_{\text{med},0} + \delta \varN_\text{med} \, , \\
    S_1(\varphi_0)e^{-\varlam_1 \varN_{\text{med},0}} = \frac{1}{2}
    \quad\implies\quad \varN_{\text{med},0} = \frac{1}{\varlam_1}\ln[2S_1(\varphi_0)] \, ,
\end{gathered}
\end{equation}
where $\varN_{\text{med},0}$ is the leading approximation from \eqref{eq:median_approx}; the correction $\delta \varN_\text{med}$ is small, and we can again expand
\begin{equation} \label{eq:median_approx_sol_2b}
\begin{gathered}
    S(\varN_\text{med}) \approx \frac{1}{2}\qty[1-\varlam_1 \delta\varN_\text{med}] + S_2(\varphi_0)e^{-\varlam_2 \varN_{\text{med},0} - \varlam_2 \delta \varN_{\text{med},0}} = \frac{1}{2} \\
    \implies\quad 
    \delta\varN_\text{med} \approx \frac{1}{\varlam_2}W_0\qty(\frac{2\varlam_2S_2(\varphi_0)}{\varlam_1}e^{-\varlam_2\varN_{\text{med},0}}) \, .
\end{gathered}
\end{equation}

Put together, the approximations $\varN_\text{med} \approx \varN_\text{cl}$ and \eqref{eq:median_approx_sol_1}--\eqref{eq:median_approx_sol_2b} cover all $\varphi_0$ values. For a given $\varphi_0$, the approximations can be checked by evaluating $S(\varN\cond\varphi_0)$ at slightly higher and lower $\varN$ values: if they lie below and above $1/2$, the approximation is good. Starting from these approximations, with slight numerical refinement, I obtained the full $\varN_\text{med}(\varphi_0)$ function depicted in Figure~\ref{fig:N_stas}.

\section{Eigensolution coefficients in an example case}
\label{sec:coefficient_list}

The following table presents the eigenvalues and eigenfunction normalization coefficients from \eqref{eq:CR_eigeneq}--\eqref{eq:CR_eigenfunctions} in the example constant-roll model with $\varphi_a=-5$, $\varphi_b=5$ for all $\varlam_n \leq 1000$. The corrections $\delta\varlam_n \equiv \varlam_n-(n-1)$ are displayed for the first 20 eigenvalues, where they are still relatively small.

\begin{center}
\begin{longtable}{lllll}
\toprule
$n$ & $\varlam_n$ & $\delta\varlam_n$ & $A_n$ & $\phantom{-}B_n$ \\
\midrule
\endhead
$1$ & $0.00$ & $3.836 \times 10^{-11}$ & $5.830 \times 10^{-6}$ & $\phantom{-}6.579 \times 10^{-6}$ \\
$2$ & $1.00$ & $1.874 \times 10^{-9}$ & $0.751$ & $-2.393 \times 10^{-9}$ \\
$3$ & $2.00$ & $4.385 \times 10^{-8}$ & $3.782 \times 10^{-8}$ & $\phantom{-}1.062$ \\
$4$ & $3.00$ & $6.527 \times 10^{-7}$ & $0.531$ & $-1.071 \times 10^{-6}$ \\
$5$ & $4.00$ & $6.929 \times 10^{-6}$ & $4.387 \times 10^{-6}$ & $\phantom{-}1.301$ \\
$6$ & $5.00$ & $5.567 \times 10^{-5}$ & $0.460$ & $-9.357 \times 10^{-5}$ \\
$7$ & $6.00$ & $3.504 \times 10^{-4}$ & $1.731 \times 10^{-4}$ & $\phantom{-}1.455$ \\
$8$ & $7.00$ & $1.763 \times 10^{-3}$ & $0.420$ & $-2.766 \times 10^{-3}$ \\
$9$ & $8.01$ & $7.178 \times 10^{-3}$ & $2.706 \times 10^{-3}$ & $\phantom{-}1.579$ \\
$10$ & $9.02$ & $2.388 \times 10^{-2}$ & $0.398$ & $-3.156 \times 10^{-2}$ \\
$11$ \hspace{0.3cm} & $10.07$ \hspace{0.3cm} & $6.576 \times 10^{-2}$ \hspace{0.3cm} & $1.732 \times 10^{-2}$ \hspace{0.3cm} & $\phantom{-}1.718$ \\
$12$ & $11.15$ & $0.153$ & $0.391$ & $-0.149$ \\
$13$ & $12.31$ & $0.308$ & $5.060 \times 10^{-2}$ & $\phantom{-}1.907$ \\
$14$ & $13.55$ & $0.548$ & $0.394$ & $-0.358$ \\
$15$ & $14.89$ & $0.885$ & $8.856 \times 10^{-2}$ & $\phantom{-}2.125$ \\
$16$ & $16.33$ & $1.325$ & $0.397$ & $-0.592$ \\
$17$ & $17.87$ & $1.869$ & $0.120$ & $\phantom{-}2.343$ \\
$18$ & $19.52$ & $2.517$ & $0.397$ & $-0.821$ \\
$19$ & $21.27$ & $3.268$ & $0.145$ & $\phantom{-}2.555$ \\
$20$ & $23.12$ & $4.121$ & $0.395$ & $-1.041$ \\
$21$ & $25.08$ & -- & $0.164$ & $\phantom{-}2.763$ \\
$22$ & $27.13$ & -- & $0.393$ & $-1.255$ \\
$23$ & $29.29$ & -- & $0.179$ & $\phantom{-}2.969$ \\
$24$ & $31.55$ & -- & $0.390$ & $-1.465$ \\
$25$ & $33.91$ & -- & $0.192$ & $\phantom{-}3.173$ \\
$26$ & $36.36$ & -- & $0.387$ & $-1.673$ \\
$27$ & $38.92$ & -- & $0.203$ & $\phantom{-}3.376$ \\
$28$ & $41.58$ & -- & $0.384$ & $-1.879$ \\
$29$ & $44.34$ & -- & $0.212$ & $\phantom{-}3.579$ \\
$30$ & $47.19$ & -- & $0.381$ & $-2.083$ \\
$31$ & $50.15$ & -- & $0.219$ & $\phantom{-}3.780$ \\
$32$ & $53.20$ & -- & $0.378$ & $-2.287$ \\
$33$ & $56.36$ & -- & $0.226$ & $\phantom{-}3.982$ \\
$34$ & $59.61$ & -- & $0.375$ & $-2.489$ \\
$35$ & $62.96$ & -- & $0.232$ & $\phantom{-}4.183$ \\
$36$ & $66.41$ & -- & $0.373$ & $-2.691$ \\
$37$ & $69.96$ & -- & $0.237$ & $\phantom{-}4.383$ \\
$38$ & $73.61$ & -- & $0.371$ & $-2.893$ \\
$39$ & $77.36$ & -- & $0.241$ & $\phantom{-}4.584$ \\
$40$ & $81.21$ & -- & $0.369$ & $-3.094$ \\
$41$ & $85.15$ & -- & $0.246$ & $\phantom{-}4.784$ \\
$42$ & $89.20$ & -- & $0.367$ & $-3.295$ \\
$43$ & $93.34$ & -- & $0.249$ & $\phantom{-}4.984$ \\
$44$ & $97.58$ & -- & $0.365$ & $-3.496$ \\
$45$ & $101.92$ & -- & $0.253$ & $\phantom{-}5.184$ \\
$46$ & $106.36$ & -- & $0.363$ & $-3.697$ \\
$47$ & $110.90$ & -- & $0.256$ & $\phantom{-}5.384$ \\
$48$ & $115.54$ & -- & $0.362$ & $-3.897$ \\
$49$ & $120.28$ & -- & $0.258$ & $\phantom{-}5.584$ \\
$50$ & $125.11$ & -- & $0.360$ & $-4.097$ \\
$51$ & $130.04$ & -- & $0.261$ & $\phantom{-}5.783$ \\
$52$ & $135.08$ & -- & $0.359$ & $-4.297$ \\
$53$ & $140.21$ & -- & $0.263$ & $\phantom{-}5.983$ \\
$54$ & $145.44$ & -- & $0.357$ & $-4.497$ \\
$55$ & $150.77$ & -- & $0.265$ & $\phantom{-}6.182$ \\
$56$ & $156.19$ & -- & $0.356$ & $-4.697$ \\
$57$ & $161.72$ & -- & $0.267$ & $\phantom{-}6.382$ \\
$58$ & $167.34$ & -- & $0.355$ & $-4.897$ \\
$59$ & $173.07$ & -- & $0.269$ & $\phantom{-}6.581$ \\
$60$ & $178.89$ & -- & $0.354$ & $-5.096$ \\
$61$ & $184.81$ & -- & $0.271$ & $\phantom{-}6.781$ \\
$62$ & $190.83$ & -- & $0.353$ & $-5.296$ \\
$63$ & $196.95$ & -- & $0.272$ & $\phantom{-}6.980$ \\
$64$ & $203.17$ & -- & $0.352$ & $-5.495$ \\
$65$ & $209.48$ & -- & $0.274$ & $\phantom{-}7.179$ \\
$66$ & $215.90$ & -- & $0.351$ & $-5.695$ \\
$67$ & $222.41$ & -- & $0.275$ & $\phantom{-}7.379$ \\
$68$ & $229.02$ & -- & $0.350$ & $-5.894$ \\
$69$ & $235.73$ & -- & $0.276$ & $\phantom{-}7.578$ \\
$70$ & $242.54$ & -- & $0.349$ & $-6.094$ \\
$71$ & $249.45$ & -- & $0.278$ & $\phantom{-}7.777$ \\
$72$ & $256.46$ & -- & $0.348$ & $-6.293$ \\
$73$ & $263.57$ & -- & $0.279$ & $\phantom{-}7.976$ \\
$74$ & $270.77$ & -- & $0.347$ & $-6.492$ \\
$75$ & $278.07$ & -- & $0.280$ & $\phantom{-}8.175$ \\
$76$ & $285.47$ & -- & $0.347$ & $-6.691$ \\
$77$ & $292.98$ & -- & $0.281$ & $\phantom{-}8.374$ \\
$78$ & $300.57$ & -- & $0.346$ & $-6.891$ \\
$79$ & $308.27$ & -- & $0.282$ & $\phantom{-}8.573$ \\
$80$ & $316.07$ & -- & $0.345$ & $-7.090$ \\
$81$ & $323.96$ & -- & $0.283$ & $\phantom{-}8.773$ \\
$82$ & $331.96$ & -- & $0.345$ & $-7.289$ \\
$83$ & $340.05$ & -- & $0.284$ & $\phantom{-}8.972$ \\
$84$ & $348.24$ & -- & $0.344$ & $-7.488$ \\
$85$ & $356.53$ & -- & $0.284$ & $\phantom{-}9.171$ \\
$86$ & $364.92$ & -- & $0.343$ & $-7.687$ \\
$87$ & $373.41$ & -- & $0.285$ & $\phantom{-}9.370$ \\
$88$ & $382.00$ & -- & $0.343$ & $-7.886$ \\
$89$ & $390.68$ & -- & $0.286$ & $\phantom{-}9.569$ \\
$90$ & $399.47$ & -- & $0.342$ & $-8.085$ \\
$91$ & $408.35$ & -- & $0.287$ & $\phantom{-}9.768$ \\
$92$ & $417.33$ & -- & $0.342$ & $-8.284$ \\
$93$ & $426.41$ & -- & $0.287$ & $\phantom{-}9.967$ \\
$94$ & $435.59$ & -- & $0.341$ & $-8.483$ \\
$95$ & $444.86$ & -- & $0.288$ & $\phantom{-}10.166$ \\
$96$ & $454.24$ & -- & $0.341$ & $-8.683$ \\
$97$ & $463.72$ & -- & $0.289$ & $\phantom{-}10.365$ \\
$98$ & $473.29$ & -- & $0.340$ & $-8.882$ \\
$99$ & $482.96$ & -- & $0.289$ & $\phantom{-}10.563$ \\
$100$ & $492.73$ & -- & $0.340$ & $-9.081$ \\
$101$ & $502.60$ & -- & $0.290$ & $\phantom{-}10.762$ \\
$102$ & $512.57$ & -- & $0.339$ & $-9.279$ \\
$103$ & $522.64$ & -- & $0.290$ & $\phantom{-}10.961$ \\
$104$ & $532.80$ & -- & $0.339$ & $-9.478$ \\
$105$ & $543.07$ & -- & $0.291$ & $\phantom{-}11.160$ \\
$106$ & $553.43$ & -- & $0.339$ & $-9.677$ \\
$107$ & $563.89$ & -- & $0.291$ & $\phantom{-}11.359$ \\
$108$ & $574.45$ & -- & $0.338$ & $-9.876$ \\
$109$ & $585.11$ & -- & $0.292$ & $\phantom{-}11.558$ \\
$110$ & $595.87$ & -- & $0.338$ & $-10.075$ \\
$111$ & $606.72$ & -- & $0.292$ & $\phantom{-}11.757$ \\
$112$ & $617.68$ & -- & $0.337$ & $-10.274$ \\
$113$ & $628.73$ & -- & $0.293$ & $\phantom{-}11.956$ \\
$114$ & $639.89$ & -- & $0.337$ & $-10.473$ \\
$115$ & $651.14$ & -- & $0.293$ & $\phantom{-}12.155$ \\
$116$ & $662.49$ & -- & $0.337$ & $-10.672$ \\
$117$ & $673.94$ & -- & $0.294$ & $\phantom{-}12.354$ \\
$118$ & $685.48$ & -- & $0.336$ & $-10.871$ \\
$119$ & $697.13$ & -- & $0.294$ & $\phantom{-}12.552$ \\
$120$ & $708.87$ & -- & $0.336$ & $-11.070$ \\
$121$ & $720.72$ & -- & $0.294$ & $\phantom{-}12.751$ \\
$122$ & $732.66$ & -- & $0.336$ & $-11.269$ \\
$123$ & $744.70$ & -- & $0.295$ & $\phantom{-}12.950$ \\
$124$ & $756.84$ & -- & $0.336$ & $-11.468$ \\
$125$ & $769.08$ & -- & $0.295$ & $\phantom{-}13.149$ \\
$126$ & $781.41$ & -- & $0.335$ & $-11.667$ \\
$127$ & $793.85$ & -- & $0.295$ & $\phantom{-}13.348$ \\
$128$ & $806.38$ & -- & $0.335$ & $-11.865$ \\
$129$ & $819.02$ & -- & $0.296$ & $\phantom{-}13.547$ \\
$130$ & $831.75$ & -- & $0.335$ & $-12.064$ \\
$131$ & $844.58$ & -- & $0.296$ & $\phantom{-}13.746$ \\
$132$ & $857.51$ & -- & $0.334$ & $-12.263$ \\
$133$ & $870.54$ & -- & $0.296$ & $\phantom{-}13.944$ \\
$134$ & $883.66$ & -- & $0.334$ & $-12.462$ \\
$135$ & $896.89$ & -- & $0.297$ & $\phantom{-}14.143$ \\
$136$ & $910.21$ & -- & $0.334$ & $-12.661$ \\
$137$ & $923.64$ & -- & $0.297$ & $\phantom{-}14.342$ \\
$138$ & $937.16$ & -- & $0.334$ & $-12.860$ \\
$139$ & $950.78$ & -- & $0.297$ & $\phantom{-}14.541$ \\
$140$ & $964.50$ & -- & $0.333$ & $-13.058$ \\
$141$ & $978.31$ & -- & $0.298$ & $\phantom{-}14.740$ \\
$142$ & $992.23$ & -- & $0.333$ & $-13.257$ \\
\bottomrule
\end{longtable}
\end{center}

\bibliographystyle{JHEP}
\bibliography{constroll}

\end{document}